\documentclass[showpacs,aps,prb,reprint,superscriptaddress]{revtex4-1}
\usepackage{amsmath}
\usepackage{amsfonts}
\usepackage{graphicx}
\usepackage{color}
\usepackage{hyperref}
\newcommand{\D}{\Delta}
\newcommand{\+}{\dagger}

\newcommand{\dn}{\downarrow}

\newcommand{\e}{\varepsilon}

\newcommand{\SC}{\mathrm{SC}}
\newcommand{\leads}{\mathrm{leads}}
\newcommand{\eire}{\mathrm{wire}}
\newcommand{\qdot}{\mathrm{dot}}
\newcommand{\dotleads}{\mathrm{dot-leads}}
\newcommand{\dotwire}{\mathrm{dot-wire}}

\newcommand{\up}{\uparrow}

\newcommand{\veck}{\mathbf{k}}

\newcommand{\meV}{{\rm meV}}
\newcommand{\mueV}{\mu{\rm eV}}

\newcommand{\comm}[2]{\left[ #1,\, #2 \right]_-}
\newcommand{\anticomm}[2]{\left[ #1,\, #2 \right]_+}
\newcommand{\gf}[3]{\langle\langle  #1 ;\, #2 \rangle\rangle_{#3}}
\newcommand{\ud}{\mathrm{d}}

\begin{document}

\title{Interaction effects on a  Majorana zero mode leaking into a quantum 
dot}
\author{David A.\ Ruiz-Tijerina}
\affiliation{Instituto de F\'{\i}sica, Universidade de S\~{a}o Paulo,
C.P.\ 66318, 05315--970 S\~{a}o Paulo, SP, Brazil}
\author{E. Vernek}
\affiliation{Instituto de F\'{i}sica de S\~ao Carlos, Universidade de
S\~ao Paulo, S\~ao Carlos, S\~ao Paulo 13560-970, Brazil}
\affiliation{Instituto de F\'isica, Universidade Federal de 
Uberl\^andia, Uberl\^andia, Minas Gerais 38400-902, Brazil.}
\author{Luis G.\ G.\ V.\ Dias da Silva}
\affiliation{Instituto de F\'{\i}sica, Universidade de S\~{a}o Paulo,
C.P.\ 66318, 05315--970 S\~{a}o Paulo, SP, Brazil}
\author{J. C. Egues}
\affiliation{Instituto de F\'{i}sica de S\~ao Carlos, Universidade de
S\~ao Paulo, S\~ao Carlos, S\~ao Paulo 13560-970, Brazil}

\date{\today}
\begin{abstract}

We have recently shown [\href{http://dx.doi.org/10.1103/PhysRevB.89.165314}{Phys. Rev. B {\bf 89}, 165314 (2014)}] that a  
non--interacting quantum dot coupled to a one--dimensional topological superconductor
and to normal leads can sustain a Majorana mode even when the dot is 
expected to be empty, \emph{i.e.}, when the  dot energy level is far above 
the Fermi level of he leads. This is due to the Majorana bound state of the 
wire leaking into the quantum dot. Here we extend this previous work by 
investigating the low--temperature quantum transport through an {\it 
interacting} quantum dot connected to source and drain leads and 
side--coupled to a topological wire. We explore the signatures of a 
Majorana zero--mode leaking into the quantum dot for a wide range of dot 
parameters, using a recursive Green's function approach. We then study the 
Kondo regime using numerical renormalization group calculations. We observe 
the interplay between the Majorana mode and the Kondo effect for different 
dot-wire coupling strengths, gate voltages and Zeeman fields. 
Our results show that a ``0.5'' conductance signature appears in  the dot 
despite the interplay between the leaked Majorana mode and the Kondo 
effect. This robust feature persists for a wide range of dot parameters, 
even when the Kondo correlations are suppressed by Zeeman fields and/or 
gate voltages. The Kondo effect, on the other hand, is suppressed by both 
Zeeman fields and gate voltages. We show that the zero--bias conductance 
as a function of the magnetic field follows a well--known universality 
curve. This can be measured experimentally, and we propose that the 
universal conductance drop followed by a persistent conductance of 
$0.5\,e^2/h$ is evidence of the presence of Majorana--Kondo physics. These 
results confirm that this ``0.5'' Majorana signature in the dot 
remains even in the presence of the Kondo effect.

\end{abstract}
\date{\today}
\maketitle
\section{introduction}
\begin{figure}[t]
\begin{center}
\includegraphics{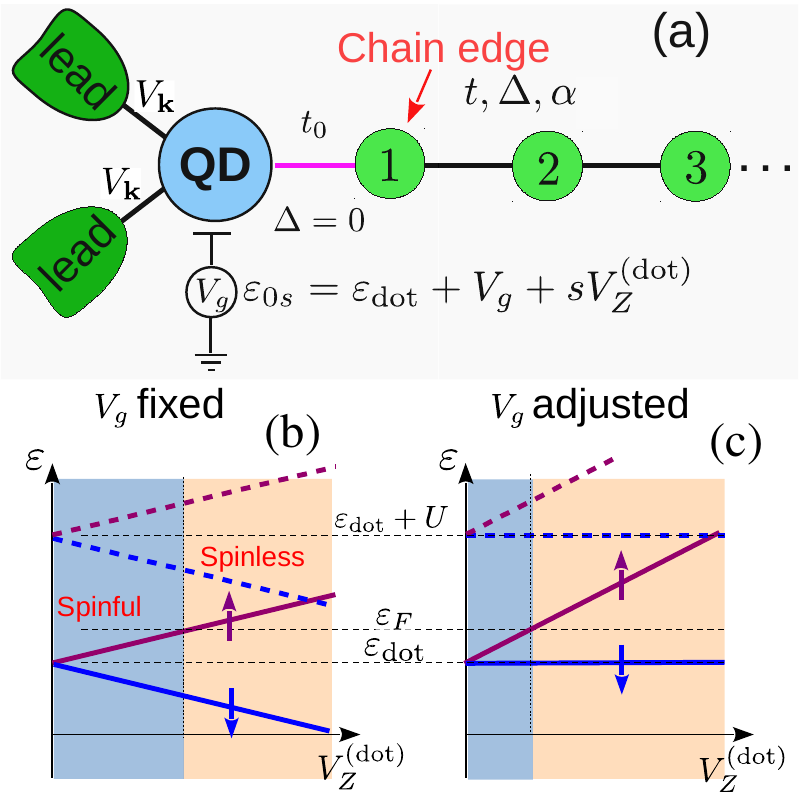}
\caption{(Color online) (a) Schematic representation of a  QD coupled to 
one end of a topological quantum wire. The quantum wire is described by a 
tight--binding chain with hopping parameter $t$, Rashba spin--orbit 
interaction $\alpha$, and induced superconducting pairing $\Delta$. 
The QD is modeled as a single orbital of energy $\e_{\text{dot}}$ with local 
Coulomb interaction $U$, coupled to metallic leads with a coupling 
$V_{\bf k}$. The energy shift in the QD level from the applied local gate 
voltage is given by $V_g$. An applied magnetic field induces a Zeeman 
splitting $V_Z$ in the wire and $V_Z^{(\text{dot})}$ in the QD, where $V_Z 
\ne V_Z^{(\text{dot})}$ because of different $g$--factors, 
$g_{\text{wire}} \ne g_{\text{dot}}$. (b) Single--particle energy level 
structure of the QD as a function of $V_Z^{(\text{dot})}$ for a fixed gate 
voltage.  (c) A spinless regime can be accessed through the application of a 
large magnetic field in the QD. For comparison with the spinful case 
 $(V_Z^{(\text{dot})}=0)$, the spin--down QD level $\varepsilon_{0,\downarrow}$ 
can be fixed at the energy $\varepsilon_{\text{dot}}$ by simultaneously applying 
a gate voltage $V_g = V_Z^{(\text{dot})}$ for positive 
$V_Z^{(\text{dot})}$.\label{fig:model}}
\end{center}
\end{figure}
The search for Majorana bound states  in condensed matter systems has 
attracted significant attention in recent years. Most of these 
investigations have focused on a geometry involving a spin--orbit coupled 
semiconducting wire with proximity--induced topological ($p$--wave) 
superconductivity, tunnel coupled to a metallic 
lead.\cite{0034-4885-75-7-076501} As it is well 
established theoretically, a finite one--dimensional (1D) topological superconductor sustains 
zero--energy mid--gap Majorana bound states at its 
ends.\cite{Phys.-Usp..44.131} Theory has predicted that these unpaired 
Majorana bound states---when the hosting superconductor is coupled to 
normal Fermi liquid leads---can give rise to a zero--bias anomaly in the 
linear conductance of the system. Mourik {\it et 
al.}\cite{Science.336.1003} were the first to report experimental 
signatures supporting this prediction in conductance measurements through 
superconductor-normal interfaces. Other theoretical and experimental 
studies\cite{NanoLetters.12.6414-6419,NaturePhysics.8.887,
PhysRevLett.109.186802,PhysRevB.87.241401,PhysRevB.86.180503, 
PhysRevB.87.024515,PhysRevB.86.155431,PhysRevB.87.060504,
PhysRevB.84.144522,NNano.2013.267} have corroborated these findings and, more importantly, 
have also pointed out a number of alternative possibilities for the 
appearance of zero--bias anomalies in transport measurements, not at all 
related to Majorana bound states (\emph{e.g.}, the Kondo effect). A review 
of these interesting possibilities is provided by Franz  in 
Ref.~\onlinecite{NatNanotechnol.8.149}.

Alternate routes to realizing Majorana  bound states have been proposed, 
involving magnetic atomic chains with spatially modulated spin textures on 
the surface of $s$--wave 
superconductors.\cite{PhysRevB.88.020407,PhysRevB.88.155420,
PhysRevLett.111.186805,PhysRevLett.111.147202,PhysRevLett.111.206802,
PhysRevB.88.180503}  In this case, a helical texture emulates the effects 
of the Zeeman plus spin--orbit fields in the earlier 
proposals,\cite{PhysRevLett.100.096407,PhysRevB.81.125318, 
PhysRevB.84.060510, PhysRevB.86.220506} thus giving rise to Majorana bound 
states. More recently, a simpler ferromagnetic configuration using 
self--assembled Fe chains on top of the strongly spin--orbit coupled 
superconductor Pb has been realized 
experimentally.\cite{yazdani_science_2014} The chain ends were probed 
locally using STM in order to directly visualize the localized modes at the 
ends of the chain. This experiment is a major step forward in the Majorana 
search, as compared to previous experimental studies, because it  provides 
the first spatially--resolved possible signature of this elusive 
state.  Note, however, that this experiment, similarly to all 
previous experiments, cannot unambiguously  associate the observed 
zero--bias  feature to the presence of Majorana 
quasiparticles.\cite{Dumitrescu}

A quantum dot (QD) attached to the extremity of a topological wire can 
also be used to locally probe the emergent Majorana end states, as proposed by 
Liu and Baranger.\cite{PhysRevB.84.201308} These authors considered a setup 
similar to Fig. 1(a) and found a conductance peak at $0.5e^2/h$ for a 
noninteracting {\it resonant} QD (\emph{i.e.}, the dot energy level 
$\varepsilon_{\rm dot}$ is aligned with the Fermi level $\varepsilon_F$ of 
the leads). Motivated by Ref.~\onlinecite{PhysRevB.84.201308}, some of us 
established in Ref.~\onlinecite{vernek_Majorana_prb_2013} that this 
feature in the QD conductance remains for a wide range of gate voltages 
$V_g$ controlling $\varepsilon_{\rm dot}$, due to the appearance of a  
resonance pinned at zero bias (\emph{i.e.}, at the Fermi level). This 
produces a conductance plateau at $0.5\,e^2/h$ spanning resonant and 
off--resonance dot level energies, far above or below $\varepsilon_F$.

We argued in Ref.~\onlinecite{vernek_Majorana_prb_2013} that this ``0.5'' 
conductance feature was quite distinct from the Kondo effect in quantum dots, as 
it would appear even for an ``empty" dot 
[$\varepsilon_{dot}(V_g)>\varepsilon_F$]. In that study, however, we had 
restricted our calculations to a non--interacting spinless model similar to that 
of Ref.~\onlinecite{PhysRevB.84.201308}. The natural question is then: How 
robust are those results, i.e., the ``0.5" plateau in the conductance, in the 
presence of the Coulomb repulsion in the dot? In particular, how is the 
conductance plateau produced solely by a Majorana mode in the noninteracting 
dot of our earlier work modified by the Coulomb interaction in the dot, 
especially in the Kondo regime? 

Recent studies have addressed the Kondo regime of a quantum dot
coupled to a topological wire and to normal Fermi liquid 
leads\cite{Golub176802,Liu-1409.3860,PhysRevB.87.241402,PhysRevB.90.195108} and 
to Luttinger leads.\cite{cheng_prx_2014} An important distinction among these 
studies is whether the topological superconducting wire is grounded or 
``floating''.\cite{footx} Our present work and that of Lee \emph{et al}. in 
Ref.~\onlinecite{PhysRevB.87.241402} consider a floating wire and, 
 with the Majorana mode coupled to the QD spin down degree of freedom, 
 obtain $G_{\downarrow}=0.5\,e^2/h$ and $G_{\uparrow}=e^2/h$, giving a 
total conductance of $1.5\,e^2/h$ in the Kondo-Majorana regime. In 
Ref.~\onlinecite{Liu-1409.3860} the authors also find $G=1.5\,e^2/h$ in a 
similar setup and further calculate the zero--frequency shot noise as an 
additional probe for the Kondo-Majorana resonance. As we discuss later, 
the robust pinning $G_{\downarrow}=0.5e^2/h$ that we find in the present 
work for the interacting case corroborates the results of our previous
work\cite{vernek_Majorana_prb_2013} and establishes their 
validity in the interacting case. None of the previous studies have 
focused on the pinning at ``0.5'' of the QD conductance as the signature of 
the leaked Majorana mode in the interacting dot or on the the influence 
of gate voltages and external magnetic fields  on the Majorana-Kondo 
physics. These are the central goals of our present work.

To address the questions in the previous paragraphs, in this paper we 
perform a thorough study of the normal-lead--QD--quantum wire 
system shown schematically in Fig.~\ref{fig:model}(a). We start off with a 
realistic model for the wire that explicitly accounts for the  Rashba 
spin--orbit interaction, proximity $s$-wave superconductivity, and a Zeeman 
term used to drive the wire from its trivial to its topological phase. We 
study this model with a recursive Green's function method, using a decoupling 
procedure known as Hubbard I 
approximation.\cite{Proc.Roy.Soc.(London).A276.238} This scheme 
allows us to describe the behavior of the QD for a wide range of parameters 
in both the trivial and topological phases of the wire. However, the 
Hubbard I approximation is  known to fail when describing the 
low--temperature regime,\cite{J.Phys.F:MetalPhys..11.2389} hence a 
nonperturbative treatment is needed. For this purpose we employ the  numerical 
renormalization group (NRG). Because treating the full quantum 
wire within the NRG is inviable, we adopt a low--energy effective 
Hamiltonian,\cite{PhysRevB.82.180516} in which the realistic wire in its 
topological phase is replaced by only two Majorana end modes. 

We find that in the interacting case, within the Hubbard I approximation, 
the pinning of the Majorana peak persists for a wide range of gate voltages 
as long as the dot is empty. However, in the single--occupancy regime of the 
dot, our mean--field calculations predict that the pinning will be  suppressed 
by Coulomb blockade when the spin up/down 
states are degenerate. By applying a large Zeeman field in the QD, we drive 
it into a  spinless regime in which Coulomb blockade does not take place 
and the non--interacting character of the dot is restored with the pinning 
appearing for both occupied and unoccupied dot as described by our results 
in Fig.~1(g) of Ref.~\onlinecite{vernek_Majorana_prb_2013}.

At low temperatures, in the absence of external Zeeman  
splitting in the dot, our NRG results show that the Majorana peak in  
fact appears also in the single--occupancy regime, in agreement with 
previous NRG studies.\cite{PhysRevB.87.241402} The ``leaked'' Majorana 
mode coexists with the Kondo effect for a QD at the particle--hole 
symmetric point, giving a  total zero--bias conductance of 
$1.5\,e^2/h$.\cite{PhysRevB.87.241402}  In this situation, we also find 
that the Majorana--QD coupling strongly enhances the Kondo temperature. In 
contrast, detuning from the particle--hole symmetric point strongly 
suppresses the Kondo peak because of an effective Zeeman splitting induced 
in the QD by the Majorana mode.\cite{PhysRevB.87.241402,cheng_prx_2014} 
However, the ``0.5'' Majorana signature is immune to the Zeeman splitting 
in the QD, so, far from the particle-hole symmetric point the 
``0.5'' conductance plateau is restored. Further, the Kondo effect can be 
progressively quenched (even at the particle--hole symmetric point)  by an 
external magnetic field. In this case the resulting zero--bias conductance 
\emph{versus} Zeeman energy follows a well--known universal curve. This 
universal behavior of the conductance for low magnetic fields and the 
persistent $0.5~e^2/h$ zero--bias conductance at large magnetic fields 
are unique pieces of evidence of the Majorana--Kondo physics in the 
hybrid QD--wire system. We emphasize that even though the phenomenology 
of interacting dots is much richer than that of their non--interacting counterparts, the 
QD Majorana resonance pinned to the Fermi level of the leads we have 
predicted in Ref.~\onlinecite{vernek_Majorana_prb_2013} appears 
in both cases.

This paper is organized as follows: In Sec.~\ref{sec:model} we introduce 
the model for the QD--topological quantum wire system. The recursive 
Green's function method is explained in Sec.~\ref{sec:recursive_gf}, and 
numerical results away from the Kondo regime are shown in Secs.\ 
\ref{sec:hubb_num} and \ref{sec:hubb_num_VZ}. We introduce a 
low--temperature effective model in Sec.~\ref{sec:effective_model} 
and numerically demonstrate its equivalence to the full model in the 
topological phase. The properties of this model are then investigated using 
the NRG method in Sec.~\ref{sec:Kondo_regime}. In 
Sec.~\ref{sec:KondoQuench} we discuss the interplay between Majorana and 
Kondo physics at low temperatures. Finally, an experimental test for this 
interplay is proposed in Sec.~\ref{sec:experimental}. Our conclusions are 
presented in Sec.~\ref{sec:conclusions}.



\section{Model}\label{sec:model}
Our model consists of a single--level QD, modeled as an  atomic site 
coupled to a \emph{finite} tight--binding chain that represents the 
one--dimensional degrees of freedom of the quantum wire 
[Fig.~\ref{fig:model}(a)]. The corresponding Hamiltonian is
\begin{equation}\label{eq:eq1}
 H=H_{\qdot}+H_{\leads}+H_{\eire}+H_{\dotleads}+H_{\dotwire},
\end{equation}
where $H_{\qdot}$ describes the isolated QD, $H_{\eire}$ is the Hamiltonian 
of the wire, and $H_{\dotwire}$ couples them at one end of the wire. The 
operator $H_{\dotleads}$ represents the tunnel coupling of the QD to source 
and drain metallic leads, which are necessary for transport measurements.
The terms describing the QD and the metallic leads are given by
\begin{subequations}\label{eq:Hfull}
\begin{equation}\label{eq:H_dot}
 H_{\qdot}=\sum_{s}\e_{0,s}c^\+_{0, s}c_{0, s}+U\,n_{0,\up}n_{0,\dn},
\end{equation}
\begin{equation}\label{eq:Hleads}
 H_\leads=\sum_{\ell \veck,s}\e_{\ell \veck, s}c^\+_{\ell \veck, s}c_{\ell
\veck, s},
\end{equation}
\begin{equation}\label{eq:Hdotleads}
H_\dotleads=\sum_{\ell \veck, s}\left(V_{\ell \veck}c^\+_{0, s}c_{\ell
\veck, s}+V^*_{\ell \veck}
c^\+_{\ell \veck,s}c_{0, s}\right),
\end{equation}
\end{subequations}
where the operator $c^\+_{0,s}$ ($c_{0,s}$)  creates (annihilates) an 
electron of spin $s$ in the QD; $\e_{0,\uparrow}=\e_{\rm 
dot}+V_g+V_Z^{({\rm dot})}$ and $\e_{0,\downarrow}=\e_{\rm 
dot}+V_g-V_Z^{({\rm dot})}$, where $\e_{\rm dot}$ is the QD energy level; 
$V_g$ represents the level shift by an applied gate voltage; and 
$V_Z^{({\rm dot})}$ is the Zeeman energy induced in the QD by an external 
magnetic field. Orbital effects from the magnetic field are neglected. The 
parameter $U$ represents the energy cost for double occupancy of the QD due 
to Coulomb repulsion, and $n_{0,s}=c^\+_{0,s} c_{0,s}$ is the QD number 
operator for spin $s$. The operator $c^\+_{\ell\veck,s}$ 
($c_{\ell\veck,s}$) creates (annihilates) an electron with spin $s$, 
momentum $\veck$, and energy $\e_{\ell\veck,s}$ in the left ($\ell=L$) or 
right ($\ell=R$) lead. The coupling constant between the QD and lead $\ell$ 
is given by $V_{\ell \veck}$, and the hybridization function is given by
\begin{equation}
\Gamma(\varepsilon) = \pi\sum_{\ell \veck,s}\left|V_{\ell\veck} 
\right|^2\delta(\varepsilon - \varepsilon_{\veck}).
\end{equation}

The terms describing the quantum wire and the QD--wire coupling are
\begin{subequations}
\begin{equation}\label{eq:H_wire}
H_{\eire}=H_0+H_R+H_{SC},
\end{equation}
\begin{equation}\label{eq:Hdotwire}
H_\dotwire=-t_0\sum_{s}\left(c^\+_{0, s}c_{1, s}+c^\+_{1, s}c_{0,
s}\right).
\end{equation}
\end{subequations}
The chain operator $c^\+_{j, s}$ ($c_{j, s}$), for $j\ge 1$, creates 
(annihilates) an electron of spin $s$ at site $j$, and the hopping constant 
$t_0$ couples the QD to the first site of the wire. The terms in 
$H_{\eire}$ are
\begin{equation}\label{eq:H_0}
\begin{split}
H_0=&\sum_{j=1, s}^N\left(-\mu+V_Z\sigma^z_{ s s}  
\right)c^\dagger_ { j s }c_{j s} \\ &-\frac{t}{2}\sum_{j=1, s}^{N-1} 
\left(c^\dagger_{j+1, s}c_{j, s } +c^\dagger_{j, s}c_{j+1, s} \right),
\end{split}
\end{equation}
where $\mu$ is the chemical potential, $\sigma^{z}$ is a Pauli matrix, $t$ 
is the nearest--neighbor hopping between the sites of the tight--binding 
chain, and $t_0$ is the hopping between the QD and the first chain site. 
The Zeeman splitting $V_Z$ from an external magnetic field (orbital effects 
are neglected in the wire as well) is assumed to be applied along the 
$z$ axis, with the wire oriented along the $x$ axis. In principle, $V_Z$ 
can be different from $V_Z^{({\rm dot})}$ because of different effective 
$g$ factors in the wire and the QD.\cite{desouza_prb_2003} The length of 
the wire is given by $aN$, where $N$ is the number of sites and $a$ the 
lattice constant.

The Rashba spin--orbit Hamiltonian is
\begin{equation}\label{eq:H_R}
H_{R}=\sum_{j=1}^{N-1} \sum_{s s^\prime}(-it_{\rm SO})c^\+_{j+1, s}
\hat z\cdot \left(\vec \sigma_{ s s^\prime}\times\hat
x\right)c_{j, s^\prime}+\text{H.c.},
\end{equation}
where $t_{\rm SO}=\sqrt{E_{\rm
SO}t}$, $E_{\rm SO}=m^*\alpha^2/2\hbar^2$, $m^*$ is the  effective electron 
mass, and $\alpha$ the Rashba spin--orbit strength in the 
wire.\cite{PhysRevB.83.094525,PhysRevB.87.024515} The proximity--induced 
$s$--wave superconductivity is described by
\begin{equation}\label{H_SC}
H_{\SC}=\Delta\sum_{j=1}^N\left(c^\+_{j,\up}c^\+_{j,\dn}+c_{j,\dn}c_
{j,\up}\right),
\end{equation}
where $\Delta$ is the (renormalized) superconducting pairing amplitude, 
assumed to be real and constant along the wire for simplicity.\cite{footy}

\section{Recursive Green's function calculation}\label{sec:recursive_gf} 
The physical quantity central to our results is the spin--resolved local 
density of states at any given site (including the QD site), defined as
\begin{equation}\label{eq:dos}
\rho_{j,s}(\e) =-\frac{1}{\pi}\text{Im}\gf{c_{j,s}}{c_{j,s}^{\dagger}}{\e},
\end{equation}
where $\gf{A}{B}{\e}$ is the retarded Green's function of operators $A$ 
and $B$ in the spectral representation. We now present an iterative 
procedure for calculating this Green's function for the Hamiltonian 
Eq.~(\ref{eq:H_wire}), using the equation of motion 
method.\cite{Sov.Phys.Usp..3.320}

Because of the spin-orbit coupling and the superconducting pairing 
in $H_{\rm wire}$ [Eq.\ (\ref{eq:H_wire})], the equation of motion for  
Eq.~\eqref{eq:dos} couples it to other types of correlation functions 
involving two creation operators. To accommodate all the needed 
Green's functions we define the matrix 
\begin{equation}\label{eq:gfmatrix-0}
\begin{split}
&{\bf G}_{i,j}(\e)=\\
&\begin{pmatrix}
\gf{ c_{i,\up}}{c^\dagger_{j, \up} }{\e}&\gf{
c_{i,\up}}{c_{j, \dn}^\+ }{\e}& \gf{ c_{i,\up}}{c_{j, \up}
}{\e} & \gf{ c_{i, \up}}{c_{j, \dn} }{\e}\\
\gf{ c_{i,\dn}}{c^\dagger_{j, \up} }{\e}& \gf{
c_{i,\dn}}{c^\+_{j, \dn} }{\e} & \gf{ c_{i,\dn}}{c_{j,
\up}}{\e} & \gf{ c_{i,\dn}}{c_{j, \dn}}{\e}\\
\gf{ c_{i,\up}^\+}{c^\dagger_{j,\up} }{\e}&\gf{
c_{i,\up}^\+}{c_{j,\dn}^\+ }{\e}&\gf{ c_{i,\up}^\+}{c_{j,\up}
}{\e}& \gf{ c^\+_{i,\up}}{c_{j,\dn} }{\e}\\
\gf{ c^\+_{i,\dn}}{c^\dagger_{j,\up} }{\e}&\gf{
c^\+_{i,\dn}}{c^\+_{j,\dn} }{\e} &\gf{
c^\+_{i,\dn}}{c_{j,\up}
}{\e} & \gf{ c^\+_{i,\dn}}{c_{j,\dn} }{\e}
\end{pmatrix}.
\end{split}
\end{equation}

We start our iterative procedure by assuming that our system has 
only the two sites $N$ and $N-1$. Applying the equation of motion to the Green's 
function ${\bf G}_{N-1,N-1}(\e)$ we obtain the Dyson equation (see 
detailed derivation in Appendix \ref{app:iterative})
\begin{equation}\label{eq:GN-1N-1}
\begin{split}
{\bf G}_{N-1,N-1}(\e)=& \,\tilde{\bf g}_{N-1,N-1}(\e) \\ &+\tilde{\bf
g}_{N-1,N-1}(\e){\bf t}{\bf G}_{N,N-1}(\e),
\end{split}
\end{equation}
whose solution is
\begin{eqnarray}\label{eq:GNN}
{\bf G}_{N-1,N-1}(\e)&=&\left[1-\tilde{\bf g}_{N-1,N-1}(\e){\bf t}
\tilde{\bf g}_{N,N}(\e){\bf t}^\+\right]^{-1}\nonumber\\
&&\times\tilde{\bf g}_{N-1,N-1}(\e).
\end{eqnarray}
In Eqs.~\eqref{eq:GN-1N-1} and \eqref{eq:GNN},
\begin{eqnarray}
\tilde{\bf 
g}_{N-1,N-1}(\e)=[1-{\bf g}_{N-1,N-1}(\e){\bf V}]^{-1}{\bf 
g}_{N-1,N-1}(\e),\nonumber\\
\end{eqnarray}
where ${\bf g}_{N-1,N-1}(\e)$ is the bare Green's function defined in 
Eq.~\eqref{eq:barelocal}, while ${\bf V}$ and $\bf t$ are the couplings 
given in 
Eqs.~\eqref{eq:v} and \eqref{eq:t}, respectively.

The Green's function \eqref{eq:GNN} describes the ``effective" 
site $N-1$ that carries all the information about the site $N$. We are 
interested, however, in the Green's function ${\bf G}_{1,1}(\e)$ that 
describes an ``effective" site $i=1$ carrying the information from all the 
other $N-1$ sites of the chain (with $N\rightarrow\infty$). To this end, a 
site $N-2$ is added to the chain and its Green's function can be 
evaluated using Eq.\ (\ref{eq:GNN}), with the substitutions 
$\textbf{G}_{N-1,N-1} \longrightarrow \textbf{G}_{N-2,N-2}$, 
$\tilde{\textbf{g}}_{N-1,N-1} \longrightarrow 
\tilde{\textbf{g}}_{N-2,N-2}$, and $\tilde{\textbf{g}}_{N,N} 
\longrightarrow \textbf{G}_{N-1,N-1}$. The correct description for the 
quantum wire is reached in the limit $N\gg 1$.
This iterative process converges to the  large--$N$ limit once the Green's 
functions of two subsequent sites $i-1$ and $i$ are identical. In our 
calculations this was strongly dependent on parameters, but the typical 
number of sites required for convergence was $N \sim 5 \times 10^4$.

Once we have reached convergence, the QD is added to the chain as site 
$i=0$, and the metallic leads are coupled to the QD. The infinite degrees  
of freedom of the lead electrons are correlated through the local Coulomb 
interaction in the QD, giving rise to an infinite hierarchy of equations of 
motion. Therefore, calculating the properties of an interacting QD within 
the Green's function formalism unavoidably requires certain approximations 
in order to truncate this system at finite order. We evaluate the Green's 
functions using a method inspired by the Hubbard I decoupling 
procedure,\cite{Proc.Roy.Soc.(London).A276.238} which allows us to close 
the 
recursive system of equations. The resulting Green's function for the QD is 
given by (see detailed derivation in Appendix \ref{app1})
\begin{widetext}
\begin{equation}\label{eq:g_dot_text}
{\bf g}_{0,0}(\e)=
\begin{bmatrix}
\tilde g_{0\up,0\up}(\e)& \frac{A_{g,\uparrow}(\e)U \langle c^\+_{0,\dn}
c_{0,\up}\rangle}{(\e- \e_{0,\up})(\e-\e_{0,\up}-U)
} &0&\frac{A_{g,\uparrow}(\e)U \langle c_{0,\dn}c_{0,\up}\rangle}{(\e-
\e_{0,\up})(\e-\e_{0,\up}-U) }\\
\frac{A_{g,\downarrow}(\e)U \langle
c^\+_{0,\up}c_{0,\dn}\rangle}{(\e-\e_{0,\dn})(\e-\e_{0,\dn}-U)
} &\tilde g_{0\dn,0\dn}(\e)&
\frac{A_{g,\downarrow}(\e)U \langle c_{0,\up}c_{0,\dn}\rangle}{(\e-
\e_{0,\dn})(\e-\e_{0,\dn}-U) }
&0\\
0&\frac{A_{h,\uparrow}(\e)U \langle c^\+_{0,\up}c^\+_{0,\dn}\rangle}
{(\e+\e_{0,\up})(\e+\e_{0,\up}+U) }&\tilde h_{0\up,0\up}(\e)&
\frac{A_{h,\uparrow}(\e)U \langle c_{0,\dn} c^\+_{0,\up}\rangle}{(\e+\e_{0,\up})
(\e+\e_{0,\up}+U)}
\\
\frac{A_{h,\downarrow}(\e)U \langle c^\+_{0,\dn}c^\+_{0,\up}\rangle}{(\e+
\e_{0,\dn})(\e+\e_{0,\dn}+U) }&0&\frac{A_{h,\downarrow}(\e)U \langle
c_{0,\up} c^\+_{0,\dn}\rangle}{(\e+\e_{0,\dn})(\e+\e_{0,\dn}+U)
}&\tilde h_{0\dn,0\dn}(\e)
\end{bmatrix},
\end{equation}
\end{widetext}
with the definitions $A_{g,s}(\e)=[1+i\Gamma g_{0s,0s}(\e)]^{-1}$,
$A_{h,s}(\e)=[1+i\Gamma h_{0s,0s}(\e)]^{-1}$, $\tilde 
g_{0s,0s}(\e)=A_{g,s}(\e) g_{0s,0s}(\e)$,
and $\tilde h_{0s,0s}(\e)=A_{h,s}(\e) h_{0s,0s}(\e)$, where
\begin{eqnarray}\label{eq:check1}
g_{0s,0s}(\e)=\frac{1-\langle n_{0,\bar
s}\rangle}{\e-\e_{0,s}}+\frac{\langle n_{0,\bar s}\rangle}{\e-\e_{0,s}-U},
\end{eqnarray}
and
\begin{eqnarray}\label{eq:check2}
h_{0s,0s}(\e)=\frac{1+\langle n_{0,\bar
s}\rangle}{\e+\e_{0,s}}-\frac{\langle n_{0,\bar s}\rangle}{\e+\e_{0,s}+U}.
\end{eqnarray}
Note that this approach requires the  self--consistent  
calculation  of the various local expectation values appearing in 
Eq.~\eqref{eq:g_dot_text}, such as the occupation of the QD $\langle 
n_{0,s} \rangle$, the spin--flip expectation value $\langle c^\dagger_{0,s} 
c_{0,\bar s} \rangle$,  and the pairing fraction $\langle c^\dagger_{0,s} 
c^\dagger_{0,\bar s} \rangle$. The last two quantities result from the 
spin--flip processes induced by the spin--orbit interaction and the 
$s$-wave pairing in the wire, respectively. Since these quantities 
are indirectly induced on the QD via its coupling to the wire, compared to 
the occupations of the dot, they are small quantities and can be 
neglected. To confirm this we have numerically evaluated their 
contributions for a wide range of parameters. The main effect of these 
terms is  to delay the convergence of the self--consistent 
calculation.

\subsection{Topological phase transition for the quantum wire} 
\label{sec:topological_transition}

For our numerical calculations we follow previous 
studies\cite{Science.336.1003,PhysRevB.87.024515} and use the following 
parameters for the quantum wire: $t=10~\meV$, $E_{\rm SO}=50~\mueV$,  
$\Delta=250~\mueV$, and $\tilde\mu=-0.01t$. As discussed in detail in 
Ref.~\onlinecite{0034-4885-75-7-076501}, the condition for the 
topological phase, where the wire sustains Majorana end states, is $|V_Z|>  
\sqrt{\tilde \mu^2+\Delta^2}\equiv |V_Z^{c}|$. For our parameters the 
topological phase transition occurs for $V_Z \approx \pm 250\,\mueV$.

In the remainder of this section, as well as in Secs.\ \ref{sec:hubb_num}  
and \ref{sec:hubb_num_VZ}, we maintain this set of parameters  
and work exclusively in the topological phase by setting 
the Zeeman splitting in the wire to $V_Z = 500\,\mueV$. The QD--leads 
hybridization is assumed constant and set to $\Gamma=1~\mueV$, and the  
QD--wire coupling is set to $t_0=40 \,\Gamma$. This choice of $t_0 > 
\Gamma$ ensures that the hybridization to the leads does not smear out any 
features of the density of states introduced by the coupling to the wire. 
The Fermi level of the leads is set as the energy reference, $\varepsilon_F 
= 0$.

\begin{figure}
\begin{center}
\includegraphics[width=\columnwidth]{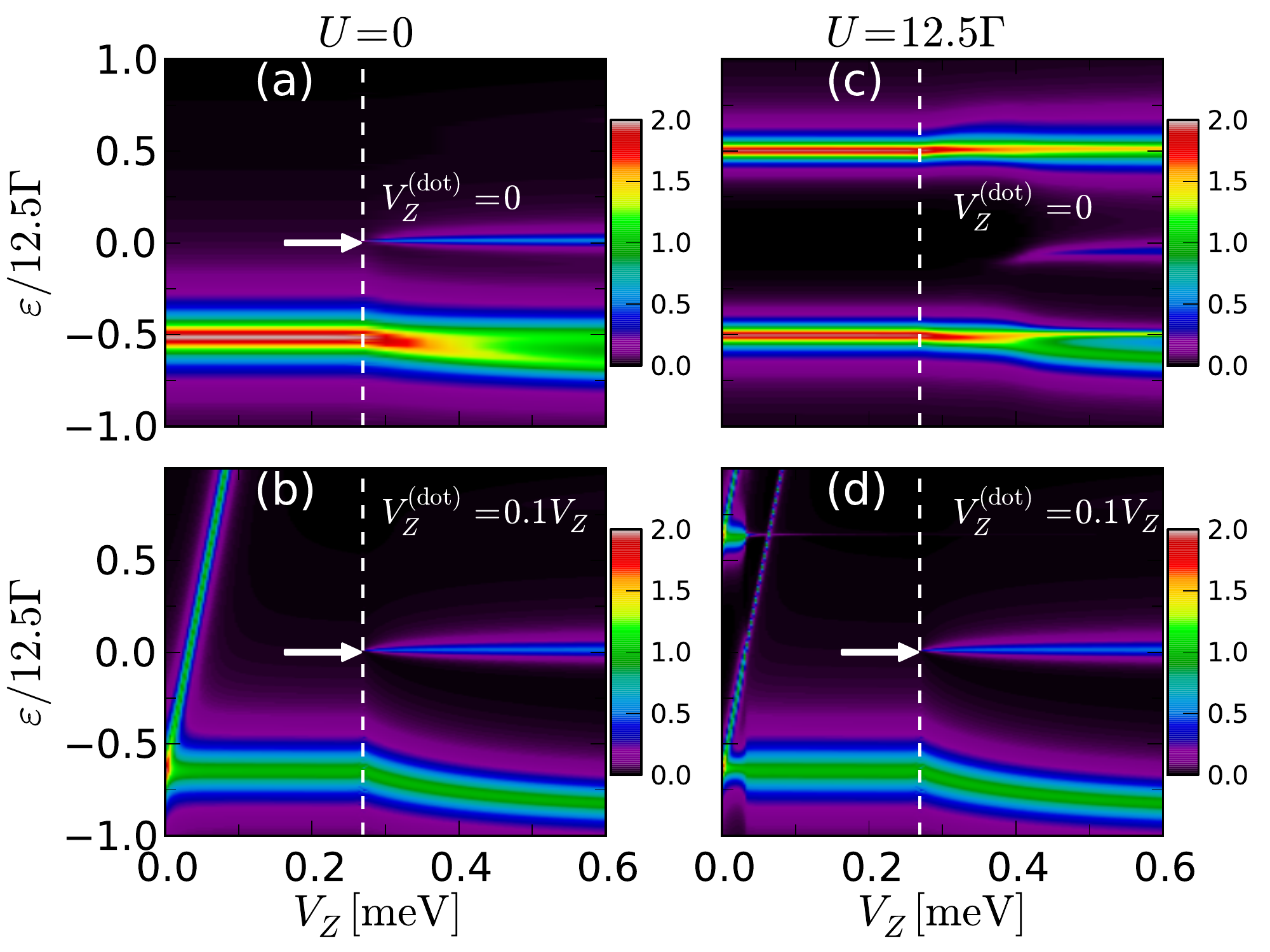}
\caption{(Color online) Density of states  at the QD site as a function of 
the Zeeman splitting in the wire, calculated using the Hubbard I method. 
The top panels [(a) and (c)] are for the QD factor $g_{\text{dot}}=0$, whereas 
the bottom panels [(b) and (d)] correspond to $g_{\text{dot}}=0.1\, 
g_{\text{wire}}$. Results for the noninteracting 
case ($U=0$) are presented in panels (a) and (b); results for the 
interacting case are shown in panels (c) and (d). These calculations are 
carried out with the QD spin--down level fixed at 
$\varepsilon_{0,\downarrow}=\varepsilon_{\text{dot}}$. For this purpose, a 
compensating gate voltage $V_g =  V_{Z}^{(\text{dot})}$ is 
applied (see the discussion in Sec.~ \ref{sec:hubb_num_VZ}). Parameters: 
$t=10~\meV$, $E_{\rm SO}=50~\mueV$, $\Delta=250~\mueV$, $\tilde\mu=-0.01t$; 
$\Gamma=1~\mueV$, $\varepsilon_{\text{dot}}=-6.25\,\Gamma$, and 
$t_0=40\,\Gamma$.\label{fig:colormap}}
\end{center}
\end{figure}

Let us begin with a general survey of the QD density of states  (DOS) 
when the wire is driven from its trivial to its topological phase, by 
increasing $V_Z>0$. Figure~\ref{fig:colormap} shows a color map of the 
total QD DOS ($\rho_\up+\rho_\dn$) \emph{versus} the energy $\varepsilon$, and 
the Zeeman energy in the wire 
$V_Z$. Henceforth we use the abbreviation $\rho_{s} \equiv 
\rho_{0,s}$ for the QD DOS. The left [Figs.\ \ref{fig:colormap}(a) and \ref{fig:colormap}(b)] and right [Figs.\ \ref{fig:colormap}(c) and \ref{fig:colormap}(d)]  
panels correspond to $U=0$ and $U=12.5\Gamma$, respectively. The top and 
bottom panels are, respectively, the DOS for $V_Z^{\rm (dot)}=0$ and 
$V_Z^{\rm dot}=0.1V_Z$. For a clear comparison among the four different 
cases we fix the lowest energy QD level---in this case 
$\varepsilon_{0,\downarrow}$, due to the positive Zeeman splitting---to an 
energy $\varepsilon_{\text{dot}}=-6.25\,\Gamma$ (see Sec.\ 
\ref{sec:hubb_num_VZ}). This is achieved with the application of a gate 
voltage $V_g=V_Z^{(\text{dot})}$, as shown in Fig.~\ref{fig:model}(c).

The general features of the DOS are as follows: In Fig.\ 
\ref{fig:colormap}(a) the colored band fixed at $\varepsilon = 
-6.25\,\Gamma$ corresponds to the spin--degenerate QD levels 
$\varepsilon_{0,s}$. When $V_Z^{(\text{dot})} = 0.1\,V_Z$ [Fig.\ 
\ref{fig:colormap}(b)] this degeneracy is broken, and the spin--up level 
$\varepsilon_{0,\uparrow}$ is seen moving to higher energies as the bright 
diagonal band on the left of the panel. The spin--down level is, as 
mentioned before, kept in place by a gate voltage $V_g = 
V_Z^{(\text{dot})}$. For $U=12.5\,\Gamma$ and $V_Z^{(\text{dot})}=0$ [Fig.\ 
\ref{fig:colormap}(c)] the spin degeneracy is restored, and so is the bright 
feature at $\varepsilon \approx -6.25\,\Gamma$. In addition, a second 
bright band appears at $\varepsilon \approx \varepsilon_{0,s}+U = 
6.25\,\Gamma$, corresponding to the doubly occupied state of the QD. When 
a large Zeeman field is introduced [Fig.\ \ref{fig:colormap}(d)] both bands 
split, shifting both the spin--up and the doubly occupied states to high 
energies, effectively eliminating them from the picture.

A sharp peak (indicated with arrows) appears at the Fermi level after the 
topological transition $V_Z>V_Z^c$  (indicated with the vertical dashed 
line) in Figs.\ \ref{fig:colormap}(a), \ref{fig:colormap}(b) and \ref{fig:colormap}(d). That is, the zero--bias signature appears 
for a non--interacting QD ($U=0$) for both a zero and a large magnetic 
field [$V_Z^{(\text{dot})} = 0$ and $V_Z^{(\text{dot})} = 0.1\,V_Z$] and 
for an interacting QD ($U=12.5\,\Gamma$) in the case of a large magnetic 
field. Note, however, that for $U=12.5\,\Gamma$ and $V_Z^{\rm (dot)}=0$ 
[Fig.~\ref{fig:colormap}(c)]  the topological phase transition appears to 
occur at higher $V_Z\approx 0.4\,\meV$. Moreover, after this apparent 
transition the central peak is strongly suppressed and shifted to negative 
energies.

The parameters used in Fig.\ \ref{fig:colormap}(c) suggest that  
these effects may be a consequence of the Coulomb blockade within the 
QD.\footnote{The results of Fig.~\ref{fig:colormap}(c) were correctly 
reproduced by the effective model Eq.\ (\ref{eq:H_eff}), with an 
appropriate choice of $\lambda$. This indicates that the apparent delayed 
transition and the appearance of a suppressed and shifted central peak is 
not related to the wire degrees of freedom, and that the topological 
transition is in fact not delayed. Then, we considered the same parameters 
of the figure, except with a small Zeeman splitting in the dot 
$V_Z^{(\text{dot})}=\Gamma$, which quenches the Kondo effect (see Fig.\ 
\ref{fig:KondoQuenchVZ}) while keeping the system in a Coulomb blockade 
($V_Z^{(\text{dot})} \ll U$). We found the appropriate value of $\lambda$ 
for the effective model to reproduce the results of the full model, and 
used it for NRG calculations. This allowed us to verify that (i) the 
shifted and suppressed central peak is an artifact of the Hubbard I 
approximation, and (ii) that the ``0.5'' peak in fact remains pinned to the 
Fermi level for that set of parameters, in agreement with our 
interpretation of the results throughout the paper.}  As mentioned above, 
when a large $V_Z^{(\text{dot})}$ is applied the spin--up and the 
doubly occupied states are pushed to high energies. For $V_Z \gtrsim 
V_Z^c$ these states no longer partake in the low--energy physics of the 
problem, and we are left with a spinless, noninteracting model. In this 
situation the zero--bias peak reappears.

As we discuss below, this peak  is associated with the formation of 
Majorana zero modes $\gamma_1$ and $\gamma_2$ at the ends of the wire. The 
mode $\gamma_1$ located close to the QD ``leaks'' into the dot, producing a 
spectral signature pinned to the Fermi level for a wide variety of QD 
parameters, in agreement with our previous 
results\cite{vernek_Majorana_prb_2013}  for a non-interacting model.
The results of Fig.~\ref{fig:colormap}(d) might suggest that the
Coulomb interaction prevents the Majorana mode from entering the QD for 
small values of $V_Z$. As we discuss in the following sections, this 
picture changes when Kondo correlations are correctly taken into account 
within the NRG approach.

\subsection{Numerical results for $V_Z^{(\text{dot})}=0$}\label{sec:hubb_num}
\begin{figure*}[t]
\begin{center}
\noindent\begin{minipage}[t]{0.98\columnwidth}
\includegraphics[width=0.95\columnwidth]{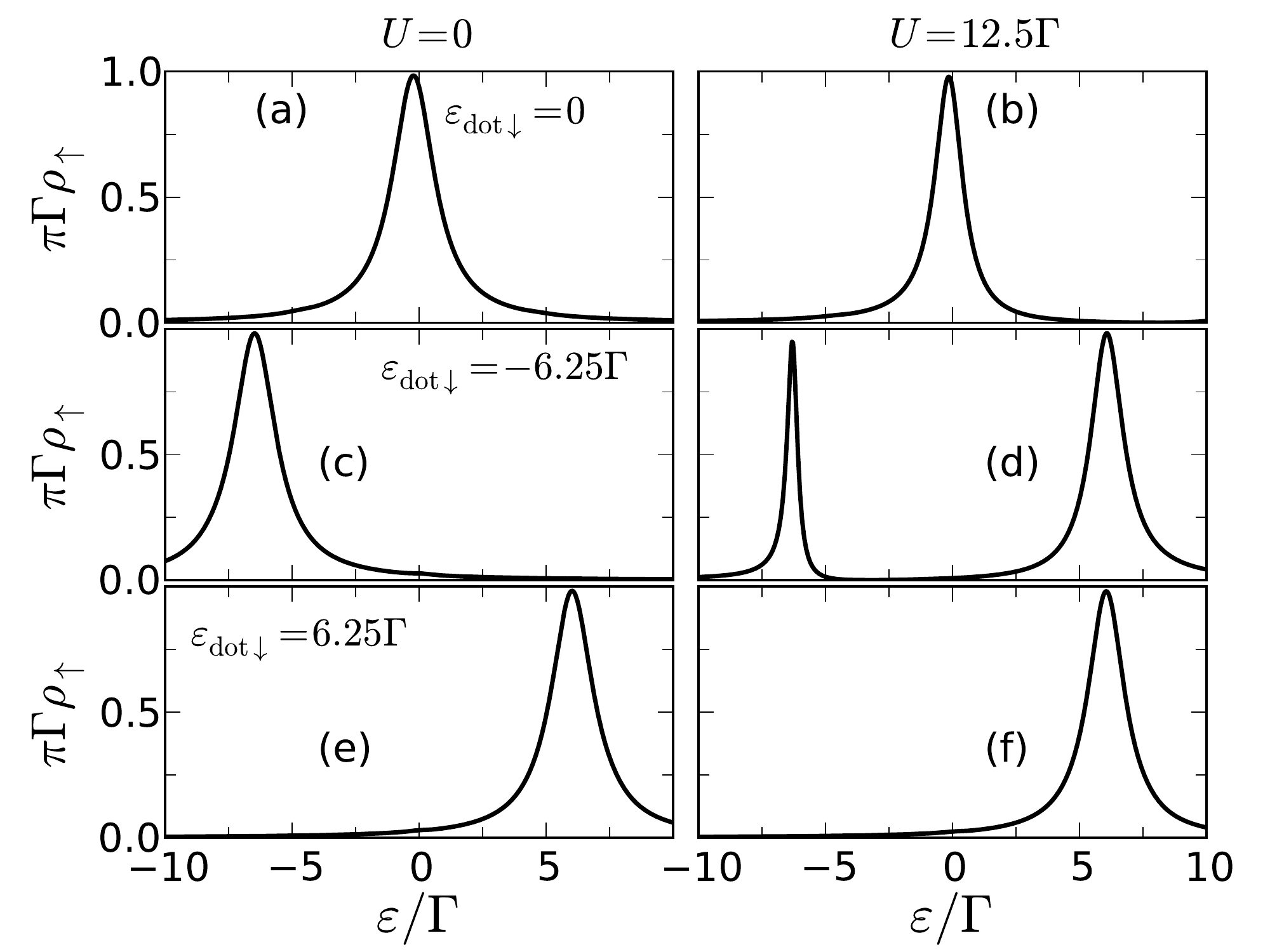}
\caption{Spin--up local density of states of the QD  for the wire in the 
topological phase, with $t=10~\meV$, $E_{\rm SO}=50~\mueV$, 
$V_Z=500~\mueV$, and $\tilde\mu=-0.01t$. QD parameters are $\Gamma=1~\mueV$ 
and $t_0=40\Gamma$.\label{fig:dos-full-up}}
\end{minipage}\qquad
\noindent\begin{minipage}[t]{0.98\columnwidth}
\includegraphics[width=0.95\columnwidth]{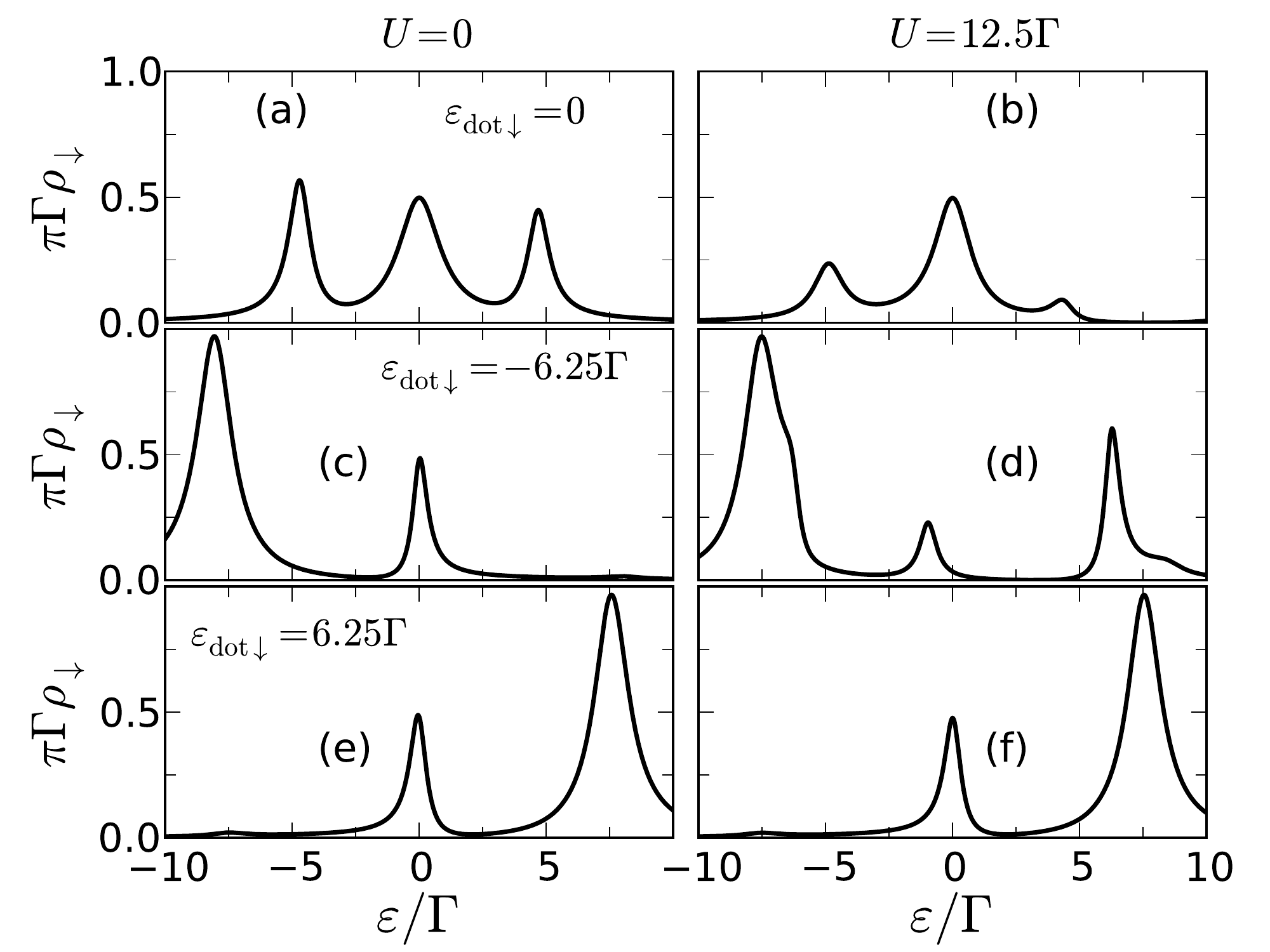}
\caption{Spin--down local density of states of the QD for the  wire in the 
topological phase, for the same parameters of Fig.\ \ref{fig:dos-full-up}. 
Note the reduced amplitude and the shift toward negative energies of the 
central peak in panel (d), for finite $U$ and 
$\varepsilon_{0,s}<0$.\label{fig:dos-full-dn}}
\end{minipage}
\end{center}
\end{figure*}

Figures \ref{fig:dos-full-up}  and \ref{fig:dos-full-dn} show the spin--up 
and spin--down local DOS at the QD site, respectively, with the wire in the 
 topological regime, and in the absence of a Zeeman splitting in the QD 
[$V_Z^{(\qdot)}=0$]. The results for an interacting ($U=12.5\Gamma$) and a 
noninteracting ($U=0$) QD are presented side by side for comparison.

The spin--up density of states in Fig.~\ref{fig:dos-full-up}  shows the 
usual structure of a QD level: In the non--interacting case there is a 
single Lorentzian peak of width $\Gamma$ and 
centered at $\varepsilon = \e_\qdot$, produced by the dot level dressed by 
the electrons of the leads. Two Hubbard bands appear in the interacting 
case, at $\varepsilon \approx \e_{\text{dot}}$ and $\varepsilon \approx  
\e_{\text{dot}} + U$ (the double occupancy excitation), but there 
are no additional features from the coupling to the quantum wire in either 
case.  This is a consequence of the large, positive Zeeman field $V_Z$ in 
the wire, which effectively decouples it from the spin--up level in the QD. 
Had we chosen a negative field $V_Z$, the spin--up level in the QD would 
decouple instead.

The signature of the Majorana zero mode forming at the end of the 
quantum wire appears in the QD spin--down density of states $\rho_{\dn}$ 
(Fig.~\ref{fig:dos-full-dn}), as an additional resonance of amplitude 
$0.5$ (in units of $1/\pi\Gamma$) pinned to the Fermi level. In the 
noninteracting case, this resonance is robust to the applied gate 
voltage [Figs.\ \ref{fig:dos-full-dn}(a), \ref{fig:dos-full-dn}(c) and \ref{fig:dos-full-dn}(e)], in agreement with our 
results for a spinless model presented in 
Ref.~\onlinecite{vernek_Majorana_prb_2013} and  also 
with Ref.~\onlinecite{PhysRevB.84.201308}. The ``0.5'' signature remains 
in the interacting case for $\varepsilon_{0,s}\ge 0$ [Figs.\  
\ref{fig:dos-full-dn}(b) and \ref{fig:dos-full-dn}(f)], and no additional features are observed 
in $\rho_{\dn}$ (apart from the two usual Hubbard bands). However, for 
$\e_{0,s}=-U/2$ [Fig.~\ref{fig:dos-full-dn}(d)], and in general for 
$\e_{0,s}<0$, with $|\e_{0,s}|\gg \Gamma$ (not shown), the central peak 
appears with a reduced amplitude ($< 0.5$) and shifted toward negative 
energies. For $V_Z < 0.4 \mueV$ [\emph{e.g.}, Fig.~\ref{fig:colormap}(c)], 
the peak can in fact be completely suppressed because of the Coulomb 
blockade in the dot. Again, we remark that this is an artifact of 
the Hubbard I approximation that is unable to correctly describe the 
ground state of the system.

\subsection{Numerical results for $V_Z^{(\text{dot})}\gg U$} 
\label{sec:hubb_num_VZ}

The only difference between the cases of Fig.~\ref{fig:dos-full-dn}(c)  
(where the ``0.5'' resonance appears) and Fig.~\ref{fig:dos-full-dn}(d) 
(where it does not) is the Coulomb interaction at the QD site. 
Indeed, the Coulomb interaction plays a role only when the dot is 
singly occupied, and there is the possibility for a second electron to hop 
into the dot (with an energy cost $U$). This is the 
situation when $\epsilon_{0,s}<\e_F$ and $\epsilon_{0,s}+U>\e_F$. The 
Coulomb blockade effect is suppressed, for instance, when the Zeeman energy 
prevents one of the spin species to hop into the dot, for example, if 
$\e_{0,\dn}<\e_F$ and $\e_{0,\up}>\e_F$ [see Figs.\ \ref{fig:model}(b) 
and \ref{fig:model}(c)]. In this case, the second electron (with spin $\up$) is prevented 
from hopping into the dot, not because of the Coulomb repulsion, but because 
of the Zeeman energy. 

\begin{figure}[b!]
\begin{center}
\includegraphics[width=\columnwidth]{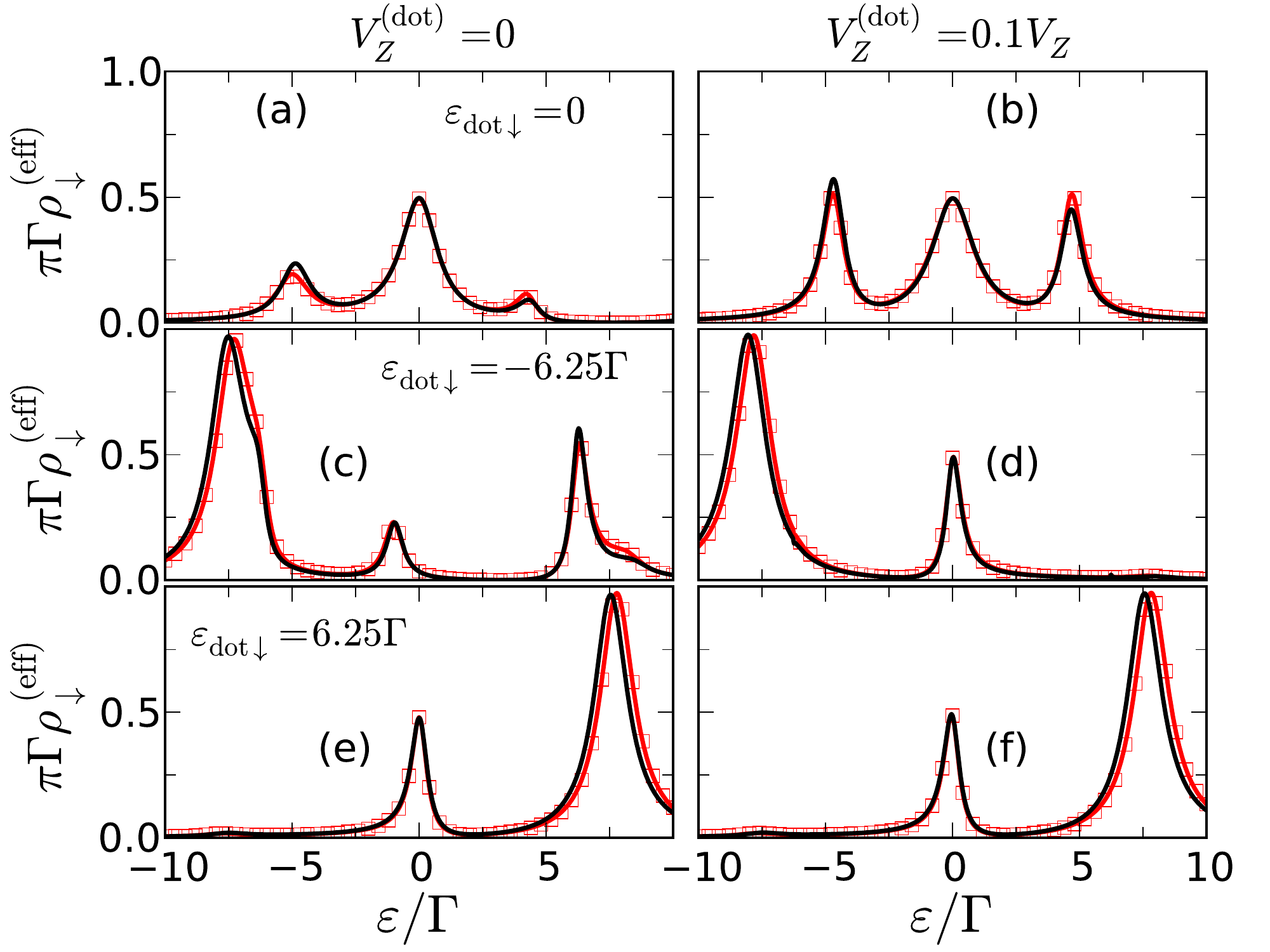}
\caption{(Color online) Spin--down local density of states at the QD  site for the 
microscopic model Eq.\ (\ref{eq:Hfull}) (solid line) and for the effective 
model Eq.\ (\ref{eq:H_eff}) (squares). The parameters are the same as in 
Fig.\ \ref{fig:dos-full-dn} but with a finite Zeeman energy $V_Z^{({\rm 
dot})}=0.1V_Z=50\mueV$. A gate voltage $V_g = V_Z^{(\text{dot})}$ is also 
applied in order to fix $\e_{0,\dn } = \varepsilon_{\text{dot}}$ for 
comparison with Fig.\ \ref{fig:dos-full-dn}.\label{fig:dot-dos-1}}
\end{center}
\end{figure}

In Fig.~\ref{fig:dot-dos-1} we shown the spin--down 
density of states with an applied Zeeman 
field $V_Z^{(\qdot)}=0.1\,V_Z$ within the QD, introduced to suppress the  
Coulomb blockade within the single--occupancy regime. The 
field raises (lowers) the spin--up (spin--down) level to $\e_{0,\uparrow} = 
\e_\qdot + V_Z^{(\qdot)}$ ($\e_{0,\downarrow} = \e_\qdot - V_Z^{(\qdot)}$), 
producing a total Zeeman splitting of $2V_Z^{(\qdot)}$. We want to 
compare the results  for the $\rho_\dn$ with finite $V_Z^{(\qdot)}$ 
with those for $V_Z^{(\qdot)}=0$ shown in Fig.~\ref{fig:dos-full-dn}. 
Since now $\varepsilon_{0,\downarrow}$ is shifted by $-V_Z^{\rm 
(dot)}$, we adjust the  gate voltage for every value of $V_Z^{(\qdot)}$ as 
$V_g = V_Z^{(\qdot)}$, so the peak of $\rho_\dn$ at  
$\e=\varepsilon_{0,\downarrow}$ appears always in the same place, 
regardless of the Zeeman energy strength in the QD. This same 
procedure, which does not pose any major experimental difficulties, was  
followed in Fig.~\ref{fig:colormap}, and is sketched in 
Figs.~\ref{fig:model}(b) and (c).

The large magnetic field and gate voltage ($V_g=V_Z^{(\qdot)}>U$)  push the 
spin--up level and the doubly occupied state to much higher energies. These 
are the bright diagonal lines seen in Figs.\ \ref{fig:colormap}(b) and \ref{fig:colormap}(d) 
moving out of the frame. This makes $\rho_{\uparrow} = 0$ in the relevant 
energy range and renders the electron--electron interaction irrelevant. At 
this point we are left with an effectively spinless model 
[Fig.~\ref{fig:model}(c)]. As expected, the large magnetic field brings 
the central peak to the Fermi level and restores its amplitude. This can 
be seen by comparing Figs.~\ref{fig:dot-dos-1}(d) and 
\ref{fig:dos-full-dn}(d). 

These results indicate that the suppression of the ``0.5''  peak in the 
case of Fig.~\ref{fig:dos-full-dn}(d) is related to the Coulomb  
blockade effect, at a Hartree level. This, however, should be 
taken with caution: As we mentioned above, the evaluation of the Green's 
function for the interacting QD requires an approximation in order to close 
the hierarchy of equations of motion. For this purpose, the Hubbard I 
method uses a mean--field approach, which by definition neglects important 
many--body correlations introduced by the Coulomb interaction.\footnote{see 
Appendix \ref{app1}, specifically Eq.~(\ref{eq:h1procedure})} Thus, the 
observed behavior of the central peak in the Coulomb blockade regime may 
well be an artifact of the method. In Sec.~\ref{sec:Kondo_regime} we 
demonstrate that this is, in fact, the case. To properly take into 
account the many--body correlations we use the NRG method for the 
low-temperature regime. However, within the NRG approach we cannot handle 
the full realistic model for the wire. We then use an effective low-energy 
model capable of describing the wire in its topological phase. This 
effective model is described next.

\section{Effective model}\label{sec:effective_model}

In this section we discuss the equivalence of the full model 
Eq.~(\ref{eq:eq1}) in the topological phase to an effective Hamiltonian in 
which only the emergent Majorana end state is directly coupled to the QD. 
This effective model has been used in the literature to describe hybrid 
QD--topological quantum wire systems, representing the wire only in terms 
of its Majorana end states.\cite{Golub176802,PhysRevB.87.241402,
cheng_prx_2014} To our knowledge, its equivalence to the full microscopic 
model has never been demonstrated.

The effective model has been employed recently in the  non--interacting QD 
limit, in which case its DOS and transport properties can be calculated 
analytically.\cite{PhysRevB.84.201308} Here, we include the Coulomb 
interaction in the QD site and use the Hubbard I approximation to close the 
infinite hyerarchy of equations of motion resulting from it. We evaluate 
its corresponding DOS $\rho_s^{\text{eff}}(\varepsilon)$ and compare it 
with the results of Secs.\ \ref{sec:hubb_num} and \ref{sec:hubb_num_VZ}, 
showing that all QD spectral features are correctly reproduced for an 
appropriate choice of the QD--Majorana parameter $\lambda$ (Fig.\ 
\ref{fig:dot-dos-1}).

The results of Sec.\ \ref{sec:hubb_num} demonstrate that only  the QD 
spin--down channel couples to the quantum wire in the topological phase 
when $V_Z > 0$. In this situation the effective model is given by
\begin{equation}\label{eq:H_eff}
\begin{split}
 H_{\rm eff}=&\sum_{s}\e_{0,s} c^\+_{0,s}c_{0,s}+U\,n_{0,\up}n_{0,\dn}+ 
\lambda\left(c_{0,\dn}-c^\+_{0,\dn} \right)\gamma_1 \\
&+ H_{\leads} + H_{\dotleads},
\end{split}
\end{equation}
where $\gamma_1$ is the operator for the Majorana bound state at the end  
of the wire, $\lambda$ is the coupling between the dot and the Majorana 
end mode\footnote{In principle, it is possible to 
express the coupling $\lambda$ in terms of $t_0$ and the parameters of the 
wire. However, as far as we know, there is no  derivation of such an 
expression. Here we justify the use of this simplified model by numerically 
finding the value of $\lambda$ in which the  results from the effective 
model \eqref{eq:H_eff} coincide with those of the full model 
\eqref{eq:H_wire}. 
We checked, for example, that for a given set of parameters of the wire 
and $t_0$, in the full model,  the equivalent $\lambda$ in the effective 
model does not depend on $\Gamma$. 
} and 
$H_{\leads}$ and $H_{\dotleads}$ are defined in Eqs.~\eqref{eq:Hleads} and 
\eqref{eq:Hdotleads}, respectively.

By using the equation of motion technique we derive the following closed
expression for the spin--down Green's function at the QD site, within the Hubbard I
approximation (Appendix \ref{app1}):
\begin{equation}\label{eq:g_simple}
\gf{ c_{0\dn}}{ c^\+_{0\dn}}{\e} = \frac{\tilde
g_{0\dn,0\dn}(\e)\left[ \e-2\lambda^2 \tilde
h_{0\dn,0\dn}(\e)\right]}{\e-2\lambda^2 \tilde
g_{0\dn,0\dn}(\e)-2\lambda^2 \tilde h_{0\dn,0\dn}(\e)}.
\end{equation}
The peak structure of the density of states $\rho^{(\rm eff)}_{\dn}(\e)$  
is the same as that from the microscopic model.  For the set of parameters  
used in Fig.~\ref{fig:dos-full-dn} we found that $\lambda=t_0/17$ 
quantitatively reproduces the results of the full chain in both the interacting 
(Hubbard-I) and the non--interacting (not shown) cases. This is presented 
in Fig.~\ref{fig:dot-dos-1}, where the density of states of the effective 
model is plotted in squares.

In the specific case of a large Zeeman field $V_Z^{(\qdot)}$ 
[Figs.\ \ref{fig:dot-dos-1}(b), \ref{fig:dot-dos-1}(d) and \ref{fig:dot-dos-1}(f)], the recovery of the central 
peak in the microscopic model is also reproduced by the effective model, as 
can be seen in Fig.~\ref{fig:dot-dos-1}(d). Moreover, the Green's function 
Eq.~(\ref{eq:g_simple}) gives further insight into the behavior of the 
``0.5'' resonance in the Coulomb blockade regime.

As the spin--up density of states $\rho_{\uparrow}^{(\text{eff})}$ vanishes 
 due to the large Zeeman splitting, so does the ground--state spin--up 
occupancy, given by
\begin{equation}
\langle n_{0,\uparrow} \rangle = \int_{-\infty}^{\infty}\ud 
\omega\,  \rho_{\uparrow}(\omega) f(\omega,\,T),
\end{equation}
with $f(\omega,\,T)$ the Fermi function. This directly relates to the 
``0.5''  resonance in the spin--down density of states. At $\e = 0$, Eq.\ 
(\ref{eq:g_simple}) can be written as
\begin{equation}
\begin{split}
\gf{ c_{0\dn}}{ c^\+_{0\dn}}{\e=0}
=\frac{1}{ [g_\dn(0)]^{-1} + [h_\dn(0)]^{-1} +2i\Gamma},
\end{split}
\end{equation}
whose density of states is a Lorentzian peak centered at the Fermi level 
only if $[g_\dn(0)]^{-1} + [h_\dn(0)]^{-1}=0$, that is, if
\begin{equation}\label{conditionn}
\frac{\e_{0,\dn}(\e_{0,\dn}+U)}{\e_{0,\dn}+U(1+\langle
n_{0,\up}\rangle )} - \frac{\e_{0,\dn}(\e_{0,\dn}+U)}{\e_{0,\dn}+U(1-\langle
n_{0,\up}\rangle )}
=0.
\end{equation}
Equation (\ref{conditionn}) is satisfied for arbitrary $\e_{0,\downarrow}$ 
only when $U=0$, or $\langle n_{0,\uparrow} \rangle=0$. The latter is 
precisely the case in Fig.~\ref{fig:dot-dos-1}(d).

The excellent agreement between the effective model and the results of 
Secs.~\ref{sec:hubb_num} and \ref{sec:hubb_num_VZ} shows that the 
effective model captures the Majorana feature both in the non--interacting and in 
the interacting regime within the Hubbard I approximation.
In Sec.\ \ref{sec:Kondo_regime} we study the Kondo regime of the 
Majorana--QD system with NRG 
method.\cite{RevModPhys.47.773,PhysRevB.21.1003,PhysRevB.21.1044}

For a typical QD--lead system, not coupled to the quantum wire,  the 
NRG method relies on the mapping of the itinerant electron degrees of 
freedom into a tight--binding chain, where each site represents a given 
energy scale. This energy scale decreases exponentially with the 
``distance'' between the QD and the chain 
site.\cite{RevModPhys.47.773,PhysRevB.21.1003} For our hybrid QD--quantum 
wire sytem, however, the gapped nature of the topological superconducting 
wire prevents us from doing this mapping, which is fundamental for treating 
the leads and the wire on equal footing. This is not a problem for the 
effective model, where the topological property of the quantum wire 
is  represented simply as a Majorana state.

\section{Kondo Regime}\label{sec:Kondo_regime}
As mentioned in Sec.\ \ref{sec:hubb_num}, the Hubbard I method makes  use 
of a mean--field approximation [Eq.\ \ref{eq:h1procedure}] which 
systematically neglects the many--body correlations introduced by the local 
Coulomb interaction within the QD. This is a good approximation at high 
temperatures, and it allows us to describe the system both in and out of 
the topological phase, as a function of all of the quantum wire parameters. 
However, for the parameters of Figs.\ \ref{fig:dos-full-up}(d) and 
\ref{fig:dos-full-dn}(d), these correlations are known to give rise to the 
Kondo effect,\cite{goldhaber_gordon_1998} which in a typical QD (not 
coupled to the quantum wire) dominates the behavior of the system below a 
characteristic temperature scale $T_K$, known as the Kondo 
temperature.\cite{abrikosov_1965} In this low--temperature regime the 
Hubbard I approximation is at a loss, and the study of the Majorana--QD 
system requires a method which can fully describe these low--energy 
correlations.

In this section we employ the NRG to study the effective Hamiltonian 
Eq.~\eqref{eq:H_eff}, which describes the relevant degrees of freedom of 
the quantum wire in terms only of the emergent Majorana zero mode at its 
end and its coupling to the QD, $\lambda$.

The NRG is a fully nonperturbative technique tailor--made to treat 
many--body  correlations in quantum impurity 
problems.\cite{PhysRevB.21.1003,PhysRevB.21.1044} It makes use of a 
logarithmic discretization of the leads' energy continuum to thoroughly 
sample the energy scales closest to the Fermi level, which are the most 
relevant for the Kondo effect.\cite{RevModPhys.47.773,RevModPhys.80.395} 
However, a well--known limitation of this discretization scheme is the 
relatively poor description of high--energy spectral features, such as 
Hubbard bands. Thus the NRG and the Hubbard I results complement each other 
for a full description of the system at hand.

The effective model can be written as a two--site interacting quantum 
impurity  with a local superconducting pairing term, coupled to metallic 
leads
\begin{equation}\label{eq:Heff_f}
\begin{split}
H_{\rm eff}=& H_{\text{dot}} + H_{\leads} + H_{\dotleads}\\
&+ \lambda\left(c^\+_{0,\dn}f_{\dn}+c^\+_{0,\dn}f^\+_{\dn} + \text{H.c.} \right),
\end{split}
\end{equation}
where $H_{\text{dot}}$, $H_{\leads}$, and $H_{\dotleads}$ are defined in 
Eq.~(\ref{eq:Hfull}), and the operator $f_{\dn}=\left(\gamma_1+i 
\gamma_2\right)/\sqrt{2}$ represents a regular fermion associated with the 
Majorana bound states in the wire. Its number operator is given by 
$n_{f,\downarrow} = f_{\downarrow}^{\dagger}f_{\downarrow}$.

One can readily see that the last term in the Hamiltonian 
Eq.~(\ref{eq:Heff_f}) does not preserve the total charge $\hat{N} = 
n_{0,\uparrow}+n_{0,\downarrow}+n_{f}$, or the total spin projection $S_z = 
(n_{0,\uparrow} - n_{0,\downarrow} - n_{f,\downarrow})/2$. However, 
defining $\hat{N}_s$ as the total number of fermions with spin index $s$, 
we see that Eq.\ (\ref{eq:Heff_f}) preserves $\hat{N}_{\uparrow}$ and the 
parity defined by the operator $\hat{P}_{\dn}\equiv(-1)^{\hat{N}_{\dn}}$. 
That is, the even or odd ($+1$ or $-1$) parity of the number of spin--down 
fermions in the Majorana--QD--leads system. This choice of quantum numbers 
considerably simplifies the NRG calculations, as noted in 
Ref.~\onlinecite{PhysRevB.87.241402}. In order to calculate the spectral 
properties of the model, we use the density matrix NRG (DM--NRG) 
method.\cite{Hofstetter:1508:2000}

\begin{figure*}[t]
\begin{center}
\begin{minipage}[t]{0.98\columnwidth}
\includegraphics[width=0.95\columnwidth]{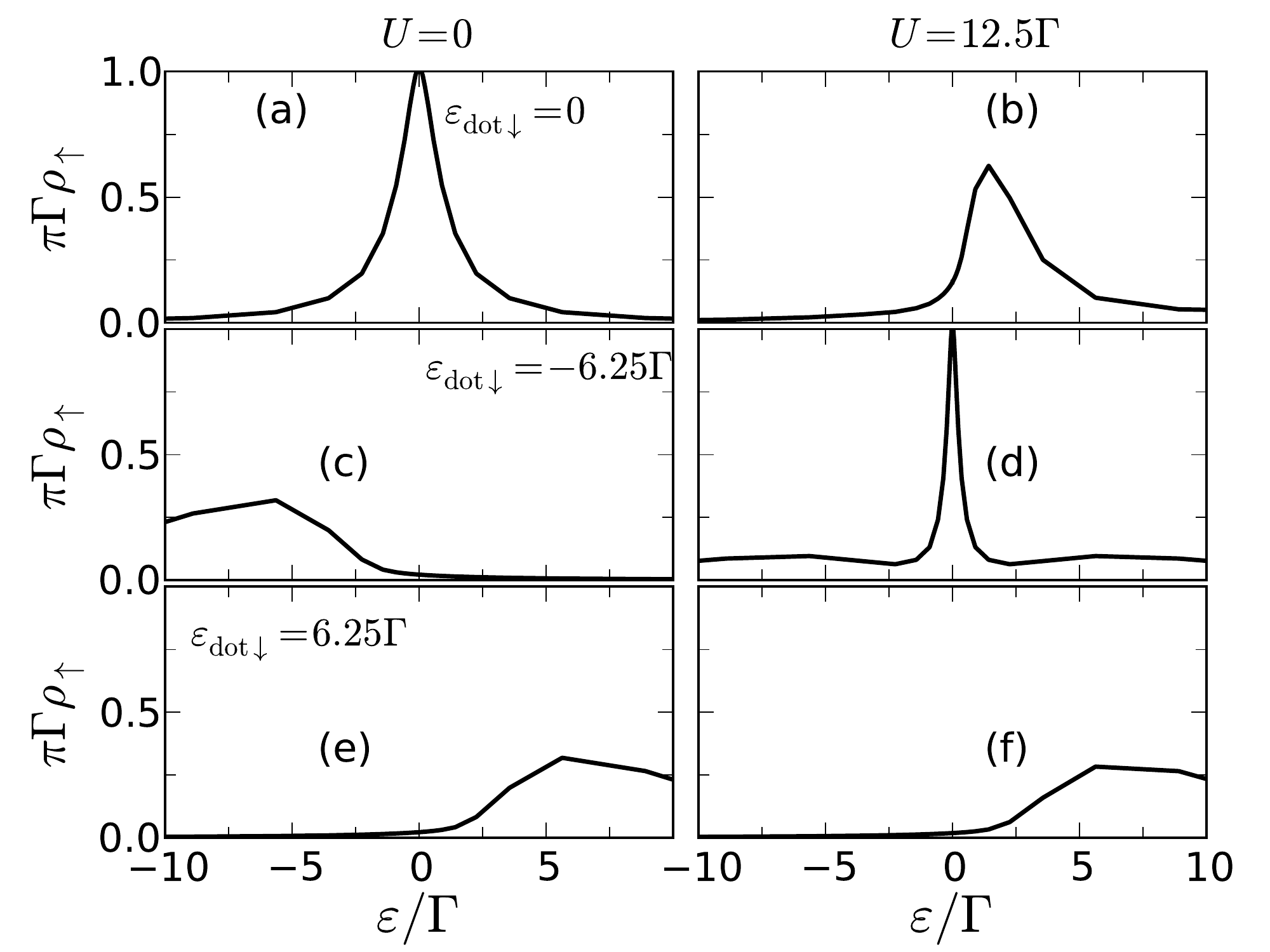}
 \caption{NRG calculations of the zero--temperature spin--up local density  
of states at the QD site, in the absence of a magnetic field 
($V_Z^{(\text{dot})}=0$). The interacting (noninteracting) case is 
presented in the right (left) panels, where the Coulomb interaction is $U = 
12.5\Gamma$ ($U = 0$). The QD level position is indicated in the panels, 
and the Majorana--QD coupling is $\lambda = 0.707\,\Gamma$. 
\label{fig:spectral_up_tall}}
\end{minipage}\qquad
\begin{minipage}[t]{0.98\columnwidth}
\includegraphics[width=0.95\columnwidth]{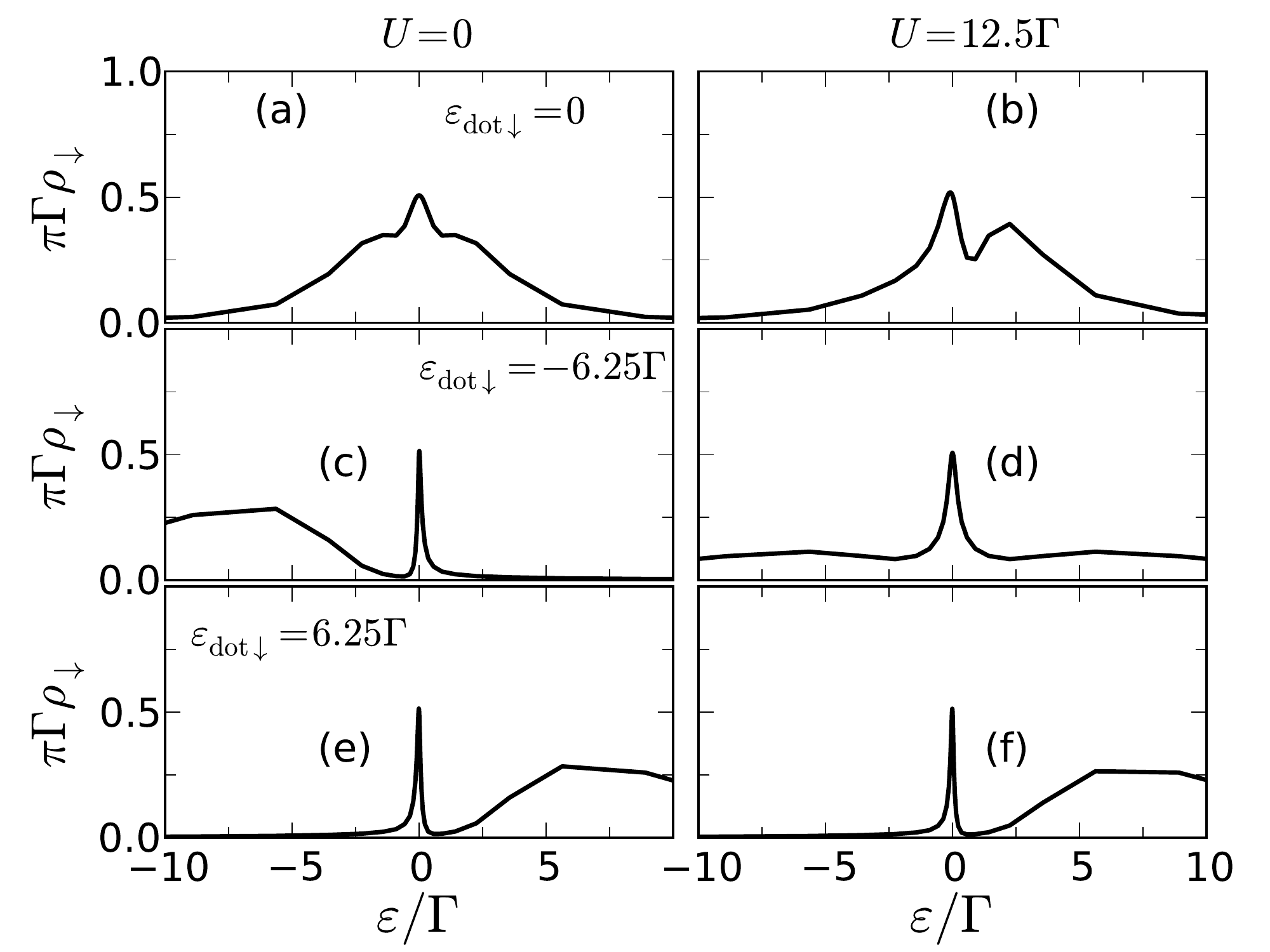}
 \caption{NRG calculations of the zero--temperature spin--down local 
density  of states at the QD site, in the absence of a magnetic field 
($V_Z^{(\text{dot})}=0$). The interacting (noninteracting) case is 
presented in the right (left) panels, where the Coulomb interaction is $U = 
12.5\Gamma$ ($U = 0$). The QD level position is indicated in the panels, 
and the Majorana--QD coupling is $\lambda = 0.707\,\Gamma$. 
\label{fig:spectral_dn_tall}}
\end{minipage}
\end{center}
\end{figure*}

\begin{figure}
\begin{center}
\includegraphics[width=\columnwidth]{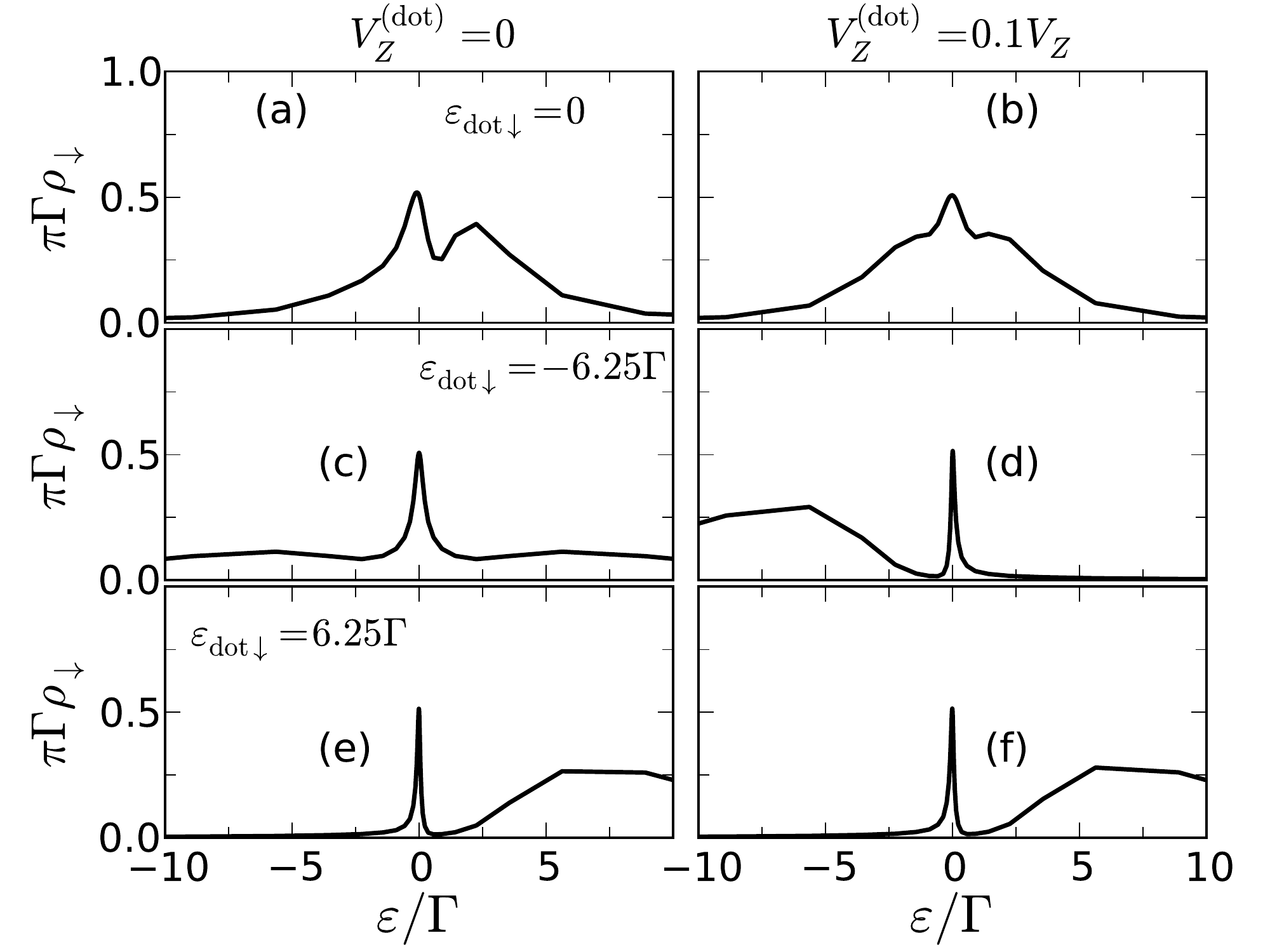}
\caption{NRG calculations of the zero--temperature spin--down local density 
 of states at the QD site, in the presence of a strong Zeeman field 
$V_Z^{(\text{dot})}$ within the QD.  Parameters: $U = 12.5\,\Gamma$, 
$\lambda = \Gamma$. \label{fig:spectral_dn_VZ} }
\end{center}
\end{figure}

The spin--resolved DOS $\rho_{\uparrow}(\varepsilon)$ and 
$\rho_{\downarrow} (\varepsilon)$ are shown in Figs.\ 
\ref{fig:spectral_up_tall} and \ref{fig:spectral_dn_tall}, respectively. 
For comparison with Figs.~\ref{fig:dos-full-up} and \ref{fig:dos-full-dn}, 
the left panels of each figure show the results for the noninteracting 
case ($U=0$), whereas the interacting case ($U>0$) is presented in the 
right panels. 

By construction, the spin--up channel has no direct coupling to the 
Majorana  degrees of freedom. As a consequence, the spin--up spectral 
density in the noninteracting case (Figs.~\ref{fig:spectral_up_tall}, 
left panels) shows only the usual Hubbard band at $\e_{0,\uparrow}$. 
Comparing to the corresponding panels of Fig.~\ref{fig:dos-full-up}, we 
can see that the position of the Hubbard band for each case is consistent 
in both calculations, although the peak is somewhat excessively broadened in the 
DM-NRG calculations, a known limitation of the broadening procedure from 
the discrete NRG spectral data.\cite{PhysRevB.84.085142}

The most important differences appear in the interacting case 
(Fig.~\ref{fig:spectral_up_tall}, right panels). For 
$\varepsilon_{0,\uparrow} = 0$, the Hubbard I approximation predicts a peak 
in the density of states at the Fermi energy ($\varepsilon = 0$), as can be 
seen in Fig.~\ref{fig:dos-full-dn}(b). This corresponds to the QD spin--up 
level, dressed by the electrons from the leads.  That is not the case for 
the NRG results, where the QD energy level appears shifted away from 
$\varepsilon = 0$ and toward positive energies  
[Fig.~\ref{fig:spectral_up_tall}(b)] due to the particle--hole asymmetry 
introduced by the Coulomb interaction in the case of  
$\varepsilon_{0,\sigma}=0$.

For $\varepsilon_{0,\uparrow} = -6.25\,\Gamma$ the QD is in the 
 single--occupancy regime, where the Kondo effect occurs at temperatures 
below $T_K$. This is signaled by the appearance of a sharp peak of 
amplitude $(\pi\Gamma)^{-1}$  and width $\sim T_K$ at the Fermi level in 
Fig.\ \ref{fig:spectral_up_tall}(d), typical of the Kondo ground state. It 
should be noted that these results correspond to $\varepsilon_{0,s} = 
-U/2$, where the QD has particle--hole symmetry. When there is some 
detuning $\delta$ from the particle--hole symmetric point, such that 
$\varepsilon_{0,s} = -U/2 + \delta$, an effective Zeeman splitting of 
strength $8|\delta|\lambda^2/U^2$ is known to arise in the QD because the 
Majorana mode couples exclusively to one spin 
channel.\cite{PhysRevB.87.241402} The Kondo effect is quenched when this 
splitting is larger than the Kondo temperature. This is in stark contrast 
to the results of Fig.~\ref{fig:dos-full-up}(d), where the Hubbard I 
approximation predicts simply a Coulomb blockade gap for all $-U < 
\varepsilon_{0,s} < 0$.

We now turn to the spin--down DOS, presented in 
Fig.~\ref{fig:spectral_dn_tall}.  The signature of the Majorana mode 
``leaking'' into the QD can be seen both in the interacting and in the 
non--interacting case, and for all values of $\varepsilon_{0,\downarrow}$. 
In the absence of interactions, the NRG calculation confirms the results 
from the Hubbard I approximation: For $\varepsilon_{0,\downarrow}=0$, shown 
in Fig.~\ref{fig:spectral_dn_tall}(a), the same three--peak structure of 
Fig.~\ref{fig:dos-full-dn}(a) is observed, albeit with wider side peaks. 
Our reasons for using a smaller value of $\lambda$ become clear in 
this case: The side bands in Fig.\ \ref{fig:spectral_dn_tall}(a) appear at 
positions\cite{PhysRevB.84.201308} $\varepsilon = \pm\lambda$. By using a 
small $\lambda$ we keep them closer to the Fermi level, where they are 
better resolved by our NRG results. In Figs.\ \ref{fig:spectral_dn_tall}(c) 
and \ref{fig:spectral_dn_tall}(e) we observe the expected Hubbard bands centered at $\varepsilon = 
\pm 6.5 \, \Gamma$, but, more importantly, the ``0.5'' peak pinned at the 
Fermi level. This is also in good agreement with the results of 
Figs.\ \ref{fig:dos-full-dn}(c) and \ref{fig:dos-full-dn}(e).

As in the case of the spin--up density of states, there are important  
differences between the results from the two methods in the interacting 
case. For $\varepsilon_{0,\downarrow}=0$, the Hubbard I results 
predict that the three--peak structure seen in the noninteracting case 
remains in the presence of the Coulomb interaction; the ``0.5'' peak 
remains intact and the amplitudes for both side bands are reduced 
[Fig.~\ref{fig:dos-full-dn}(b)]. In contrast, the NRG results of 
Fig.~\ref{fig:spectral_dn_tall}(b) demonstrate that the side bands are 
shifted to positive energies, as in the case of the spin up level in 
Fig.~\ref{fig:spectral_up_tall}(b). The left side band is strongly reduced 
and mixes with the tail of the ``0.5'' central peak, which remains pinned 
to the Fermi level in the presence of the Coulomb interaction.

Figure \ref{fig:spectral_dn_tall}(d) shows that the ``0.5'' peak persists  
even in the single--occupancy regime ($\varepsilon_{0,\downarrow} = 
-6.5\,\Gamma$), where in a typical QD (in the absence of the wire) the 
Kondo peak would be expected. We emphasize that the NRG method is 
particularly accurate at energies close to the Fermi level, and that it 
correctly describes this signature of the Majorana mode. This important 
result has also been found by Lee \emph{et al}.;\cite{PhysRevB.87.241402} 
it demonstrates that the Majorana ground state dominates over the Kondo 
effect at zero temperature and that the signature is robust to the effects 
of the Coulomb interaction in the QD. This was recently discussed in 
Ref.~\onlinecite{cheng_prx_2014},  using an analytical renormalization 
group analysis of a similar system in the weak QD--Majorana coupling limit. 
There it was suggested that a new low--energy Majorana fixed point emerges, 
which dominates over the usual (Kondo) strong--coupling fixed point. This 
picture is certainly supported by our results.

For completeness, we evaluate also the spin--down density of states in the 
limit  of a large Zeeman field in the interacting QD [$V_Z^{(\text{dot})} 
\gg U>0$] for comparison with the results presented in 
Fig.~\ref{fig:dot-dos-1}. As discussed in Sec.~\ref{sec:hubb_num_VZ}, the 
combination of the positive Zeeman splitting and the gate voltage holding 
the spin--down level in place raises the spin--up level to high energies. 
This effectively freezes the spin--up and the double occupancy states, 
restoring the noninteracting picture and eliminating the possibility for 
the Kondo effect. This is shown in the right panels of Fig. 
\ref{fig:spectral_dn_VZ}. Figures \ref{fig:spectral_dn_VZ}(a), \ref{fig:spectral_dn_VZ}(c) and \ref{fig:spectral_dn_VZ}(e) show the same results 
as the corresponding panels of Fig.~\ref{fig:spectral_dn_tall} for 
side by side comparison. As expected, the large magnetic field restores 
the results for a noninteracting QD, presented in the left panels of 
Fig.~\ref{fig:spectral_dn_tall}. This is consistent with the Hubbard I 
results of Fig.~\ref{fig:dot-dos-1}.

\section{Separating the Kondo--Majorana ground 
state}\label{sec:KondoQuench}
\begin{figure}
\includegraphics[width=\columnwidth]{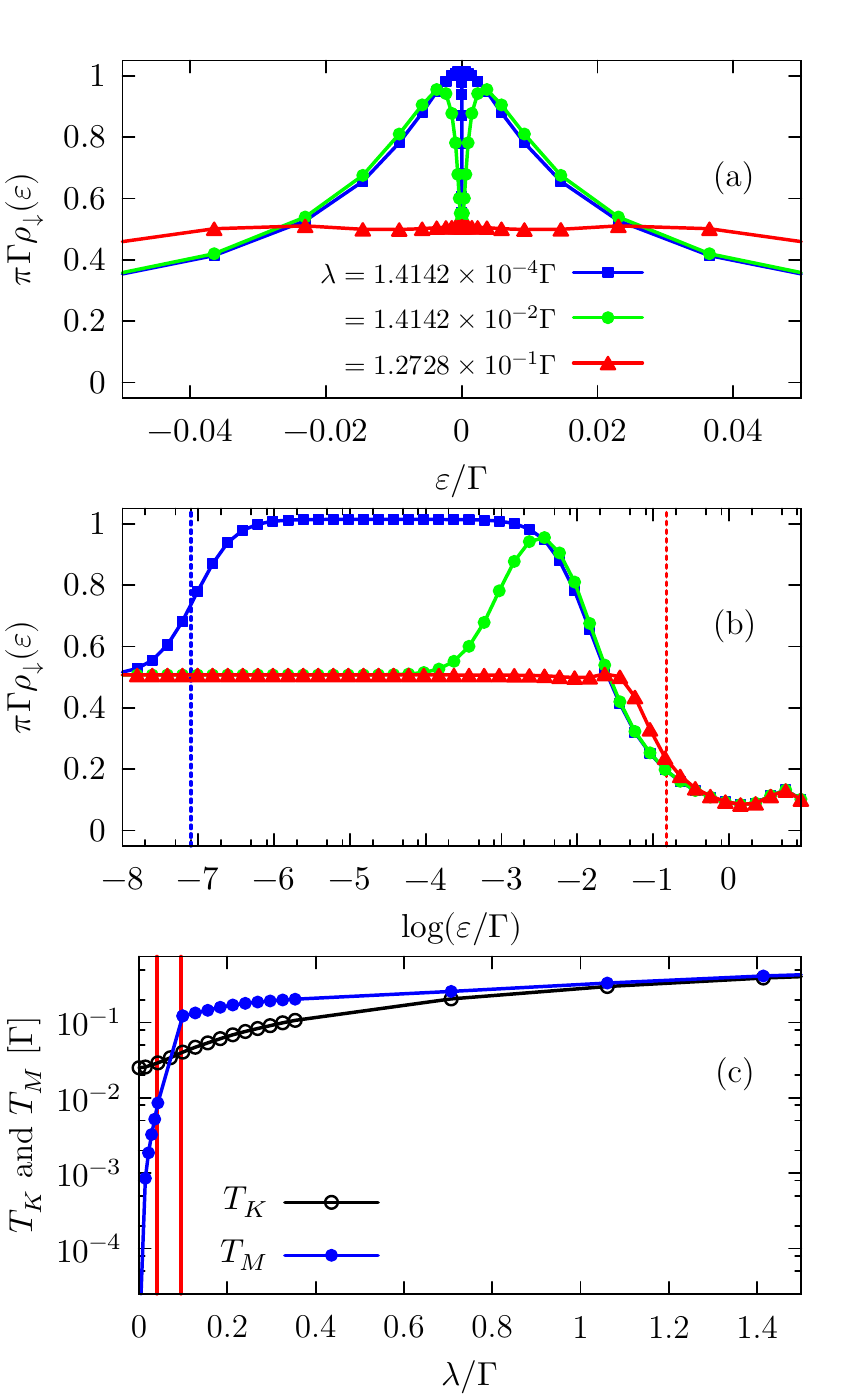}
\caption{(Color online) Extracting the Majorana energy scale $T_M$  from 
the spin--down density of states, for three values of $\lambda$. The 
spin--down DOS is shown close to the Fermi level in panel (a). In panel (b) 
it is shown with the energy axis in logarithmic scale. For $T_M \gg T_K$ 
(triangles), $T_M$ is given by the width at half--maximum of the ``0.5'' 
peak. In that case, $T_M \sim 10^{-1}\,\Gamma$. When $T_M \ll T_K$ 
(squares), $T_M$ is given by the width of the dip that indicates the onset 
of the Majorana--dominated regime, in this case giving $T_M \sim 
10^{-7}\,\Gamma$. In the Kondo--Majorana crossover (circles) $T_M$ cannot 
be clearly distinguished. (c) $T_K$ and $T_M$ in units of $\Gamma$, as 
functions of $\lambda$. The Kondo temperature was extracted from the 
spin--up density of states as the width at half--maximum of the Kondo 
peak.\label{fig:width}}
\end{figure}

The results of Figs.~\ref{fig:spectral_up_tall} and  
\ref{fig:spectral_dn_tall} have established that the DOS of the interacting 
QD near particle--hole symmetry features mixed Kondo and Majorana 
signatures. According to a recent study,\cite{cheng_prx_2014} the QD 
spin--down channel is strongly entangled with the Majorana mode and the 
lead electrons through the conservation of the parity 
$\hat{P}_{\downarrow}$ defined in Sec.~\ref{sec:Kondo_regime}. As a 
consequence, the ``0.5'' peak is strongly renormalized by the QD--lead 
hybridization $\Gamma$. This was demonstrated in 
Ref.~\onlinecite{PhysRevB.87.241402}, where the Majorana energy scale was 
shown to depend on the hydridization as $\lambda/\Gamma$.

The QD spin--up channel, on the other hand, exhibits Kondo correlations  
which arise through virtual spin--flip processes between the lead electrons 
and the QD spin--up and spin--down levels. The persistence of the Kondo 
effect suggests that, despite its entanglement with the Majorana mode, the 
spin--down degree of freedom of the singly occupied QD takes part in these 
processes. It follows that the Kondo temperature---the width of the 
zero--bias peak in $\rho_{\uparrow}$---must be renormalized by the 
Majorana--QD coupling $\lambda$.

In Fig.~\ref{fig:width} we present the dependence of the Kondo ($T_K$) and 
 Majorana ($T_M$) energy scales on $\lambda$, as extracted numerically from 
the density of states. The Kondo temperature was calculated as the width at 
half--maximum of the zero--bias peak in $\rho_{\uparrow}$. As for $T_M$, 
the process was somewhat subtler and requires some clarification.

Consider the top curve (squares) in Fig.~\ref{fig:width}(a), where 
$\lambda \ll \Gamma$. In this case the Kondo temperature is $T_K \sim 
10^{-2}\,\Gamma$, and the Majorana scale is $T_M \sim 10^{-7}\,\Gamma$. The 
former is obtained from $\rho_{\uparrow}$ (not shown), but for this value 
of $\lambda$ it can also be seen in $\rho_{\downarrow}$, as shown in 
Fig.~\ref{fig:width}(b): With $\varepsilon$ presented in a logarithmic 
scale, the positive--energy half of the Kondo peak looks simply as a climb 
to the $(\pi\Gamma)^{-1}$ plateau, going from right to left (higher to 
lower energies). This climb corresponds to the crossover to the (Kondo) 
strong--coupling fixed point, and its width at half--maximum gives $T_K$. 
Then, there is a drop to a $0.5(\pi\Gamma)^{-1}$ plateau, which represents 
a dip in the middle of the Kondo peak [Fig.~\ref{fig:width}(a)]. This 
corresponds to the crossover to the Majorana fixed point, and $T_M$ (marked 
by the vertical line on the left) is given by the energy half--way into the 
drop.

Consider now the bottom curve (triangles) in Fig.~\ref{fig:width}(a), 
 where $\lambda \gtrsim \Gamma$ and $T_M \gg T_K$. In this situation the 
crossover to the Majorana fixed point is a climb instead of a drop, and 
$T_M \sim 10^{-1}\,\Gamma$ can be obtained as the width of the ``0.5'' peak 
at half--maximum [right vertical line in Fig.~\ref{fig:width}(b)]. For 
intermediate cases such as that of the middle curve (circles), where $T_M 
\sim T_K$, the crossover to the Majorana fixed point mixes with the 
crossover to the Kondo fixed point, and we are unable to clearly resolve 
it.

The full dependence of $T_M$ and $T_K$ on $\lambda$ is shown in 
Fig.~\ref{fig:width}(c). The energy scale $T_M$ (solid circles) is seen to 
sharply increase until exceeding the Kondo temperature for $\lambda < 
0.1\,\Gamma$. It then enters a stage of much slower growth, until 
matching---somewhat counterintuitively---the value of $T_K$ for $\lambda 
\gtrsim \Gamma$. The two curves continue together for larger values of 
$\lambda$.

The Kondo temperature (empty circles) is smallest for $\lambda = 0$,  where 
it depends exclusively on the QD parameters, and is significantly enhanced 
by increasing the Majorana--QD coupling. This can be explained in terms of 
the spin--flip processes that give rise to the Kondo effect: In the absence 
of the Majorana mode, the spin of the singly occupied QD is flipped by 
virtual charge excitations to zero and double occupancy. The coupling to 
the Majorana mode introduces additional spin--flip processes that 
renormalize the Kondo scale, accompanied by parity exchange between 
the Majorana mode and the lead electrons.\footnote{This becomes apparent 
when the Hamiltonian Eq.~(\ref{eq:Heff_f}) is projected onto an effective 
Kondo Hamiltonian using a Schrieffer--Wolff\cite{schrieffer_wolff_PR_1966} 
transformation. We do not present this here, as it has been shown 
previously in Refs.~[\onlinecite{PhysRevB.87.241402}] (supplementary 
material) and [\onlinecite{cheng_prx_2014}].}

\section{Experimental test for the presence of a Majorana zero mode} 
\label{sec:experimental}

We now address the problem of distinguishing the Majorana  zero mode from 
the Kondo resonance through transport measurements on the QD. The 
zero--bias conductance through the QD is given by the Landauer--type 
formula\cite{PhysRevLett.68.2512}
\begin{equation}\label{eq:landauer}
G(T) = \pi\Gamma G_0\sum_{\sigma}\int \ud \omega \,  
\rho_{\sigma}(\omega)\left(-\frac{\partial f(\omega,\,T)}{\partial \omega} 
\right), 
\end{equation}
with $f(\omega,\,T)$ the Fermi function and $G_0=e^2/h$ the quantum of  
conductance. At low temperatures ($T \lesssim T_K$) Eq.~(\ref{eq:landauer}) 
can be approximated by
\begin{equation}
G = \pi\Gamma G_0\sum_{\sigma}\rho_{\sigma}(0),
\end{equation}
which is directly proportional to the sum of the spectral density 
amplitudes of both spin channels at the Fermi level. For a 
singly--occupied, particle--hole symmetric QD, and in the absence of a 
Zeeman splitting $V_Z^{(\text{dot})}$, Figs.~\ref{fig:spectral_up_tall}(d) 
and \ref{fig:spectral_dn_tall}(d) predict a low--temperature conductance 
$G=1.5\,G_0$.
\begin{figure}
\includegraphics[width=\columnwidth]{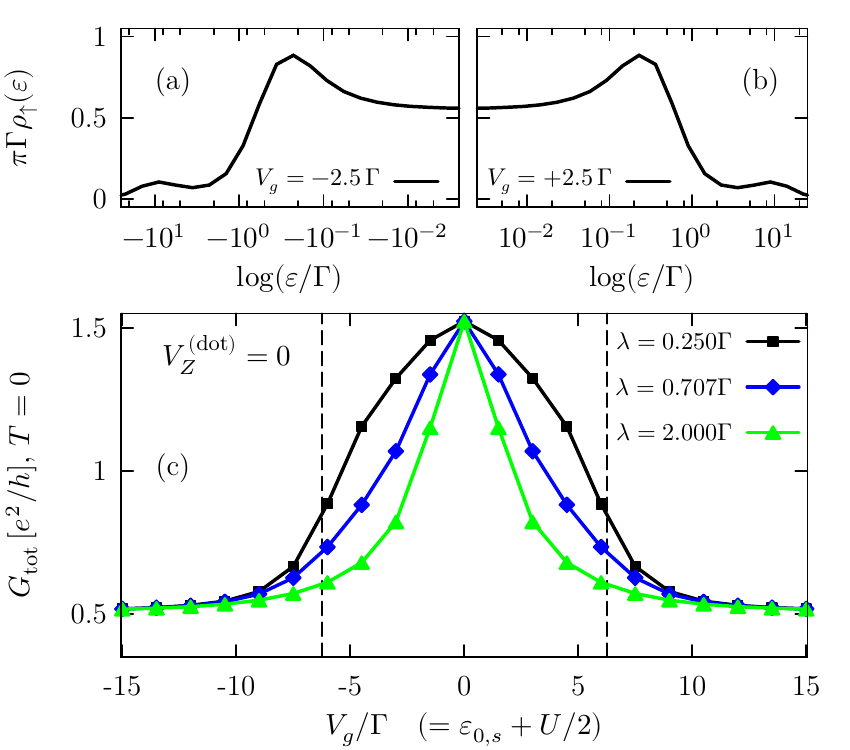}
\caption{(Color online) Conductance effects of the detuning from 
particle--hole  symmetry produced by a gate voltage $V_g$. For 
$|V_Z^{({\rm M})}| \lesssim T_K$, the spin--up DOS at the Fermi level drops 
as the Kondo peak  is shifted to positive (negative) values for $V_g > 0$ 
($V_g < 0$), due to  the Majorana--induced Zeeman splitting 
$V_Z^{(\text{M})}$. This is shown in panels (a) and (b), respectively. The 
Kondo effect finally disappears for $|V_g| \gtrsim U/2$ due to the charge 
fluctuations of the mixed--valence regime. The conductance effects of the 
Kondo quench are shown in panel (c). The enhanced conductance at zero 
detuning ($V_g=0$) comes from both the ``0.5'' and the Kondo peaks, whereas 
after the suppression of the Kondo effect the persistent $0.5\,G_0$ 
conductance comes only from the ``0.5'' peak. Parameters: 
$\varepsilon_{\text{dot}} = -U/2 = -6.25\,\Gamma$; $V_Z^{(\text{dot})}=0$.}
\label{fig:KondoQuenchVg}
\end{figure}
While establishing such a specific value in a transport experiment is far 
from trivial, a much simpler test for the presence of the Majorana mode 
can be carried out by quenching the Kondo effect using gate voltages or 
magnetic fields. A conductance drop will be observed as the Kondo resonance 
disappears but the conductance signature of $G=0.5\,G_0$ from the Majorana 
mode remains. The ratio of the conductance before and after the Kondo 
quench can be used as an indicator of the Majorana physics.

The Kondo effect occurs when the QD is close to particle--hole symmetry.  
When the QD level is detuned from the symmetric point by a gate voltage 
$V_g$, an effective Zeeman splitting arises from the spin--symmetry 
breaking induced by the Majorana mode, which couples exclusively to the 
spin--down degree of freedom. For a small gate voltage $V_g \ll |\varepsilon_{\text{dot}}|$ this splitting is given to first order in 
$V_g$ as\cite{PhysRevB.87.241402}
\begin{equation}\label{eff_Zeeman}
V_Z^{(\text{M})} = V_g\frac{8\lambda^2}{U^2}.
\end{equation}
In terms of the minimal effective model, this field appears because only 
the spin--down electron of the dot is coupled to the Majorana mode. This 
spin asymmetry is explicitly introduced in the underlying microscopic 
tight--binding model by the magnetic field in the nanowire. Virtual 
processes involving  the Majorana mode, the QD, and the band electrons lower 
the energy of the spin--down level through holelike excitations and that 
of the spin--up level indirectly through particle--like excitations. Thus, 
within the low--energy effective model, the Majorana mode gives rise to a 
Zeeman splitting within the QD when the dot is not particle--hole symmetric 
($\varepsilon_{0,s} \ne -U/2$).

The effective Zeeman splitting \eqref{eff_Zeeman} has an important 
effect on the spin--up spectral density shown in  
Figs.\ \ref{fig:KondoQuenchVg}(a) and \ref{fig:KondoQuenchVg}(b) for $|V_Z^{(\text{M})}| \lesssim 
T_K$. As this Zeeman splitting increases, the amplitude of the spin--up 
density of states at the Fermi level is reduced, and the Kondo effect is 
quenched; this reduces the low--temperature conductance, as shown in 
Fig.~\ref{fig:KondoQuenchVg}(c). In contrast and more importantly, 
the $0.5$ peak  of the spin--down density of states remains pinned at the 
Fermi level (not shown). For $\lambda = 2\,\Gamma$ (triangles) the 
effective Zeeman splitting $V_Z^{(\text{M})}$ strongly suppresses the 
Kondo 
effect, even for small $V_g$, well within the single--occupancy regime. In 
the case of $\lambda = 0.25\,\Gamma$ the splitting $V_Z^{(\text{M})}$ is 
weaker, and the Kondo effect is ultimately quenched when the QD enters the 
mixed--valence regime---that is, when its charge begins fluctuating between 
single and double occupancy ($V_g \approx -U/2$) or between single 
and zero occupancy ($V_g \approx U/2$)---as indicated by the vertical 
dashed lines (see Appendix \ref{app:splitting}). Note, however, that the spin--down contribution to the 
conductance is fixed at $0.5(\pi\Gamma)^{-1}$ for all values of $V_g$ due 
to the robustness of the ``0.5'' peak.  

\begin{figure}
\includegraphics[width=\columnwidth]{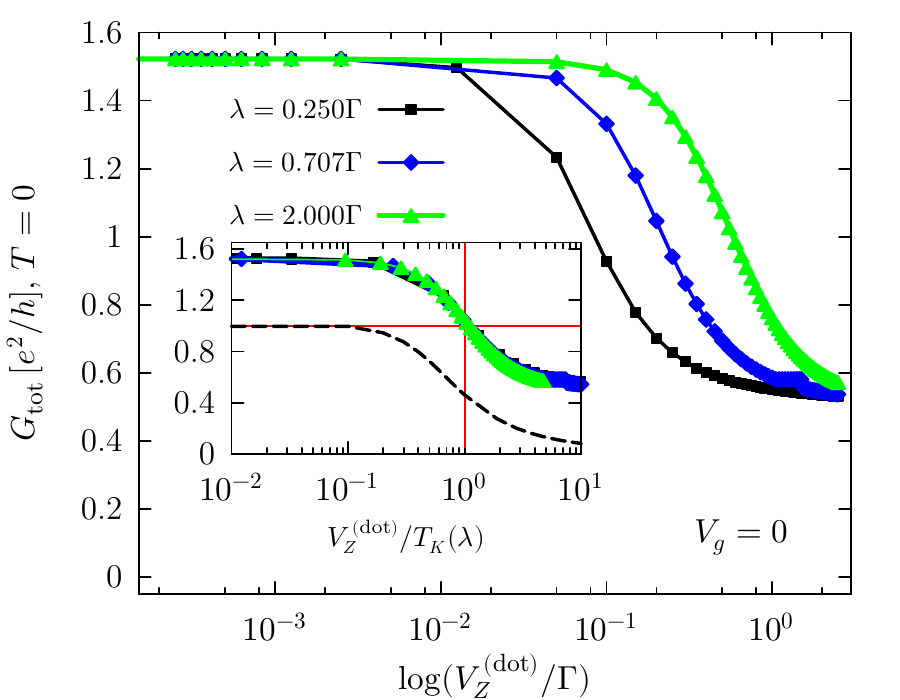}
\caption{(Color online) Low--temperature zero--bias conductance of  the QD 
as a function of the Zeeman splitting 
$V_Z^{(\text{dot})}$. Inset: Universality curve resulting from the 
rescaling $V_Z^{(\text{dot})}/T_K(\lambda)$. $T_K(\lambda)$ was obtained 
from Fig.~\ref{fig:width}. The dashed curve corresponds to the spin--up 
conductance for $\lambda=0$, and is given as a reference. Parameters: 
$\varepsilon_{\text{dot}}=-U/2=-6.25\,\Gamma$.}
\label{fig:KondoQuenchVZ}
\end{figure}

Perhaps more illuminating are the effects of an induced magnetic field,   
which breaks the spin degeneracy that is indispensable for the formation of 
the Kondo ground state. A Zeeman field of strength $V_Z^{(\text{dot})} 
\gtrsim T_K$ will suppress the Kondo zero--bias peak, and hence reduce the 
low--temperature conductance. This conductance suppression follows a 
well--known universality curve when the Zeeman splitting is rescaled by the Kondo 
temperature.\cite{Hewson-Kondo,PhysRevLett.85.1504} This is shown in 
Fig.~\ref{fig:KondoQuenchVZ}, where three different values of the coupling 
$\lambda$ are considered. The same behavior is observed for all three 
cases: a conductance plateau of $G=1.5\,G_0$ for $V_Z^{(\text{dot})}\ll 
T_K$, followed by a monotonic decrease with $V_Z^{(\text{dot})}$ until 
reaching another plateau of $G = 0.5\,G_0$ when $V_Z^{(\text{dot})} \gg 
T_K$, due solely to the Majorana mode. Moreover, the conductance \emph{versus} 
$V_Z^{(\text{dot})}$ curve for a Kondo QD is known to be 
universal.\cite{J.Phys.:Condens.Matter..} The 
three curves in Fig.~\ref{fig:KondoQuenchVZ} collapse onto a 
single universal curve by rescaling the Zeeman splitting in terms of the 
Kondo temperature $T_K(\lambda)$ corresponding to each value of $\lambda$, 
which we obtained from Fig.~\ref{fig:width}. The experimental observation 
of this curve provides certainty that the Kondo effect was present in the 
QD, and that the applied magnetic field has eliminated it from the picture, 
leaving only the Majorana zero--bias peak.

All of the results presented in this section can be measured 
experimentally  and may provide a method for detecting the emergence of 
the Majorana zero mode at the end of the topological quantum wire. When 
the QD is near its particle--hole symmetric point and the Kondo signature 
mixes with the Majorana peak, a finite zero--bias conductance can be 
measured. The Kondo effect can then be removed with the introduction of a 
gate--compensated magnetic field, leaving a finite zero--bias conductance 
coming from the ``0.5'' Majorana peak. The ``1.5 to 0.5'' ratio of the 
initial and final conductances is a clear sign of Majorana physics. 
The 
Kondo quench can be verified from the universal behavior of the 
conductance as a function of the applied magnetic field.

We emphasize the importance of fixing the QD spin--down energy level by  
means of a gate voltage, since both relevant energy scales for transport in 
the QD, $T_K$ and $T_M$, are strongly dependent on its value. It is also 
important that these experiments be carried out at a sufficiently low 
temperature, $T \ll T_K,\,T_M$. It is desirable that the Majorana energy 
scale be of the order of the Kondo temperature ($T_M \sim T_K$), because in 
the case of $T_M \ll T_K$ it may be difficult to resolve the ``0.5'' dip in 
the Kondo resonance at the Fermi level from the conductance measurements, 
especially if it is somewhat broadened by thermal effects. Using the 
wire parameters from Sec.~\ref{sec:hubb_num}, and with $\Gamma \sim 
100\,\mu\text{eV}$, the Kondo and Majorana temperature scales are 
approximately $300\,\text{mK}$.

\section{Conclusions}\label{sec:conclusions}

We have studied the low--temperature transport properties of a hybrid  
QD--topological quantum wire system, using a model that explicitly includes 
Rashba spin--orbit coupling and induced $s$--wave superconductivity in the 
quantum wire, and the local Coulomb interaction within the QD. Using 
recursive Green's function calculations, we showed that only one of the QD 
spin degrees of freedom  couples to the Majorana zero mode emerging at the 
end of the wire, whereas the other fully decouples. This is signaled by a 
zero--bias peak in the spin--resolved conductance of the QD, which is 
robust to the application of arbitrarily large gate voltages and Zeeman 
fields.

Through numerical calculations, we show that the low-energy physics 
of this full model can be captured by a minimally coupled effective 
Hamiltonian for both a noninteracting and an interacting quantum dot.
These models have been extensively used in the literature to 
describe the interaction between a quantum impurity and a Majorana fermion.

The effective model was investigated using the numerical renormalization 
group.  We studied the interacting regime of the QD, where the Kondo effect 
appears and the mean--field Green's function calculations are no longer 
valid. Our results show that the Majorana signature persists and 
suggest a QD ground state where Majorana and Kondo physics coexist.

Finally, we proposed a method for identifying the interplay between 
Majorana  and Kondo physics in the system. The QD zero--bias conductance 
should be measured close to particle--hole symmetry, where the 
Majorana and the Kondo physics coexist; this value is taken as a reference. 
The Kondo effect should be quenched by a Zeeman field in the QD, and the 
field--dependent conductance measured. For a large Zeeman splitting the 
conductance will be determined by the ``0.5'' peak, giving a value of 
$0.5\,e^2/h$---a third of the reference conductance---corresponding only to 
the Majorana signature. The quenching of the Kondo effect can be verified 
from the universality properties of the conductance \emph{versus} Zeeman field 
curves.

\begin{acknowledgments}
All authors acknowledge support from the Brazilian agencies CNPq, CAPES,  
FAPESP, and PRP/USP within the Research Support Center Initiative (NAP 
Q-NANO). E.V., J.C.E.\ and D.A.R.T.\ thank P. H. 
Penteado for helpful discussions of our results. E.V.\ and J.C.E.\ 
thank the Kavli Institute for Theoretical Physics  
(Santa Barbara) for the hospitality during the Spintronics program/2013 
where part of this work was carried out. E.V.\ acknowledges support from 
the Brazilian agency FAPEMIG. Finally, D.A.R.T.\ thanks J. D. Leal--Ruiz for inspiration and encouragement 
during the preparation of this article.
\end{acknowledgments}

\appendix

\section{Iterative equations for the Green's function of the quantum wire} 
\label{app:iterative}
We make use of the spectral representation
of the retarded Green's function\cite{Sov.Phys.Usp..3.320}
\begin{eqnarray}
 \gf{A}{B}{\e} \equiv -i\int e^{i\e\tau} \Theta(\tau) \langle  
\anticomm{A(\tau)}{B(0)}\rangle d\tau,
\end{eqnarray}
where $A(\tau)$ and $B(\tau)$ are the operators $A$ and $B$ in the 
Heisenberg  picture, and $A$ and $B$ represent any combination of fermion 
operators in the Hamiltonian. The (anti--)commutator is written as 
$[A,\,B]_{\pm}=AB\pm BA$, and $\langle \cdots \rangle$ is the thermodynamic 
average at finite temperature, or the ground--state expectation value in 
the case of zero temperature. From the standard equation of
motion technique we have the recursion relation\cite{Sov.Phys.Usp..3.320}
\begin{eqnarray}\label{EOM}
 \e \gf{A}{B}{\e} = \langle \anticomm{A}{B}\rangle+\gf{ \comm{A}{H}}{B}{\e}.
\end{eqnarray}
In order to show how we obtained the iterative procedure in a pedagogical
fashion, let us start by calculating
the local Green's function for the site $N-1$. We assume for the time being that the
wire has only two other sites: the sites $N-2$ and  $N$. This will allow us to
see how the structure of the iterative procedure for
arbitrary $N$ emerges. Given that we are ultimately interested in the local
density of states
\begin{eqnarray}
\rho_{j,s}(\e)= -\frac{1}{\pi}{\rm Im} \gf{
c_{j,s}}{c^\+_{j,s}}{\e},
\end{eqnarray}
we will start by calculating the Green's function $\gf{c_{j,s}}{
c_{j,s}}{\e}$. 
Using Eq.~\eqref{EOM} we
can write the expressions for $\gf{c_{j,\up}}{
c_{j,s^\prime}}{\e}$ and  $\gf{ c_{j,\dn}}{
c_{j,s^\prime}}{\e}$ as
\begin{equation}\label{G_up}
\begin{split}
&(\e-\e_{N-1,\up})\gf{ c_{N-1, \up}}{c^\dagger_{N-1, s^\prime}}{\e}\\
&\quad =\delta_{s^\prime\up} - \frac{t}{2} \gf{ c_{N,\up}
}{c^\dagger_{N-1,s^\prime}}{\e}+\Delta \gf{ c^\dagger_{N-1, \dn}} 
{c^\dagger_{N-1, s^\prime}}{\e}\\
&\qquad -t_{\rm SO}\gf{ c_{N,\dn} }{c^\dagger_{N-1,s^\prime}}{\e},
\end{split}
\end{equation}
and
\begin{equation}\label{G_down}
\begin{split}
&(\e-\e_{N-1,\dn})\gf{ c_{N-1, \dn}}{c^\dagger_{N-1, s^\prime}
}{\e}\\
&\quad =\delta_{s^\prime\dn}- \frac{t}{2}\gf{ c_{N,\dn}
}{c^\dagger_{N-1, s^\prime}}{\e} -\Delta\gf{ c^\dagger_{N-1, \up}} 
{c^\dagger_{N-1, s^\prime}}{\e}\\
&\qquad +t_{\rm SO}\gf{ c_{N,\up} }{c^\dagger_{N-1, s^\prime}}{\e}.
\end{split}
\end{equation}
In Eqs.~\eqref{G_up} and  \eqref{G_down} we have defined
$\e_{j,s}=-\mu+\sigma^z_{ss}V_Z$  in order to simplify the
notation. The second term on the right--hand side of each equation
describes simply the hopping between adjacent sites of the wire. The
third term describes a hopping between adjacent sites, accompanied by a
spin flip due to the Rashba spin--orbit coupling. Finally,  the fourth term
pairs electrons of opposite spin within a given site due to the $s$--wave
superconductivity. We now need to calculate  the equations
of motion for these additional correlation functions. For instance, for the
pairing correlation function we obtain
\begin{equation}
\begin{split}
&(\e+\e_{N-1,\dn})\gf{ c^\dagger_{N-1, \dn}}{ c^\dagger_{N-1,
s^\prime}}{\e}\\
&\quad = \Delta\gf{ c_{N-1, \up} }{c^\dagger_{N-1,
s^\prime}}{\e}+\frac{t}{2}\gf{ c^\dagger_{N,\dn} }{c^\dagger_{N-1,s^\prime}}{\e}\\
&\qquad -t_{\rm SO}\gf{ c^\dagger_{N, \up} }{c^\dagger_{N-1, s^\prime}}{\e},
\end{split}
\end{equation}
and
\begin{equation}
\begin{split}
&(\e+\e_{N-1,\up})\gf{ c^\dagger_{N-1, \up}}{c^\dagger_{N-1,
s^\prime}}{\e} \\
&\quad = -\Delta\gf{ c_{N-1, 
\dn}}{c^\dagger_{N-1,s^\prime}}{\e}+\frac{t}{2} \gf{ c^\dagger_{N,\up}
}{c^\dagger_{N-1, s^\prime}}{\e}\\
&\qquad+t_{\rm SO}\gf{ c^\dagger_{N, \dn}}{c^\dagger_{N-1, s^\prime}}{\e}.
\end{split}
\end{equation}

From the structure of the equations above it becomes clear that we
can define a matrix for each chain site, which contains all of the 
correlation  functions at that site:
\begin{equation}\label{eq:gfmatrix}
\begin{split}
&{\bf G}_{i,j}(\e)=\\
&\begin{pmatrix}
\gf{ c_{i,\up}}{c^\dagger_{j, \up} }{\e}&\gf{
c_{i,\up}}{c_{j, \dn}^\+ }{\e}& \gf{ c_{i,\up}}{c_{j, \up}
}{\e} & \gf{ c_{i, \up}}{c_{j, \dn} }{\e}\\
\gf{ c_{i,\dn}}{c^\dagger_{j, \up} }{\e}& \gf{
c_{i,\dn}}{c^\+_{j, \dn} }{\e} & \gf{ c_{i,\dn}}{c_{j,
\up}}{\e} & \gf{ c_{i,\dn}}{c_{j, \dn}}{\e}\\
\gf{ c_{i,\up}^\+}{c^\dagger_{j,\up} }{\e}&\gf{
c_{i,\up}^\+}{c_{j,\dn}^\+ }{\e}&\gf{ c_{i,\up}^\+}{c_{j,\up}
}{\e}& \gf{ c^\+_{i,\up}}{c_{j,\dn} }{\e}\\
\gf{ c^\+_{i,\dn}}{c^\dagger_{j,\up} }{\e}&\gf{
c^\+_{i,\dn}}{c^\+_{j,\dn} }{\e} &\gf{
c^\+_{i,\dn}}{c_{j,\up}
}{\e} & \gf{ c^\+_{i,\dn}}{c_{j,\dn} }{\e}
\end{pmatrix}.
\end{split}
\end{equation}
With this notation, the system of equations can be written as
\begin{equation}\label{GGG}
\begin{split}
 {\bf G}_{N-1,N-1}(\e)=&\,\, {\bf g}_{N-1,N-1}(\e)\\
 &+{\bf g}_{N-1,N-1}(\e){\bf V}{\bf G}_{N-1,N-1}(\e)\\
 &+{\bf g}_{N-1,N-1}(\e){\bf t}{\bf G}_{N,N-1}(\e),
\end{split}
\end{equation}
where we have defined the bare local Green's function for a generic site,
\begin{equation}\label{eq:barelocal}
{\bf g}_{j,j}(\e)=
\begin{pmatrix}
\frac{1}{\e-\e_\up}&0&0&0\\
0&\frac{1}{\e-\e_\dn}&0&0\\
0&0&\frac{1}{\e+\e_\up}&0\\
0&0&0&\frac{1}{\e+\e_\dn}
\end{pmatrix},
\end{equation}
and the matrices
\begin{subequations}\label{eq:vandt}
\begin{equation}\label{eq:v}
 {\bf V}=
\begin{pmatrix}
0&0&0&\D\\
0&0&-\D&0\\
0&-\D&0&0\\
\D&0&0&0
\end{pmatrix}, 
\end{equation}
\begin{equation}\label{eq:t}
{\bf t}=
\begin{pmatrix}
\frac{-t}{2}&-t_{\rm SO}& 0 &0\\
-t_{\rm SO}&\frac{-t}{2}&0&0\\
0&0&\frac{-t}{2}&t_{\rm SO}\\
0&0&-t_{\rm SO}&\frac{-t}{2}
\end{pmatrix},
\end{equation}
\end{subequations}
which, respectively,  pair the electrons in each site of the chain and
allow for the electrons to hop between adjacent sites, either
preserving the spin projection or flipping it. Moreover, we can write Eq.~\eqref{GGG}
in the more compact Dyson equation form
\begin{equation}\label{GGG-1}
\begin{split}
{\bf G}_{N-1,N-1}(\e)=& \,\tilde{\bf g}_{N-1,N-1}(\e) \\ &+\tilde{\bf
g}_{N-1,N-1}(\e){\bf t}{\bf G}_{N,N-1}(\e),
\end{split}
\end{equation}
with the definition
\begin{eqnarray}\label{eq:gtilde}
\tilde{\bf g}_{j,j}(\e)=(1-{\bf V})^{-1}{\bf g}_{j,j}(\e).
\end{eqnarray}
Repeating these steps for the nonlocal Green's function  ${\bf 
G}_{N,N-1}(\e)$,  we obtain
\begin{equation}\label{GGGG}
{\bf G}_{N,N-1}(\e)=\tilde {\bf g}_{N,N}(\e) \hat{\bf t}^\+{\bf
G}_{N-1,N-1}(\e).
\end{equation}
Finally, substituting Eq.~\eqref{GGGG} into \eqref{GGG-1} we find
\begin{equation}\label{GNNapp}
\begin{split}
{\bf G}_{N-1,N-1}(\e)=&\left[1-\tilde{\bf g}_{N-1,N-1}(\e){\bf t}
\tilde{\bf g}_{N,N}(\e){\bf t}^\+\right]^{-1} \\ 
&\times \tilde{\bf g}_{N-1,N-1}(\e).
\end{split}
\end{equation}
Equation \eqref{GNNapp} establishes the iterative procedure, in which for
site $N-1$ we simply replace  $\tilde{\bf g}_{N-1,N-1}(\e)$ by
${\bf G}_{N-1,N-1}(\e)$, which was calculated in the very first iteration.
This procedure can be repeated $N$ times in order to obtain the
full numerical Green's function at one end of the wire. For very
large $N$, ${\bf G}_{1,1}(\e)$ will be indistinguishable from ${\bf
G}_{2,2}(\e)$; at that point the semi-infinite chain limit will have been  
reached.

\section{The Green's function of the quantum dot \label{app1}}
For the derivation of the local Green's function of the QD, we assume that  
the QD is symmetrically coupled to the right and left terminals and 
replace them by a symmetrized band with a coupling $\tilde
V_{\veck}=\sqrt{2}V_{\ell\veck}$. The total hybridization function for the  
symmetric band is given by $\Gamma(\omega) = \pi\sum_{\veck}\, 
|\tilde{V}_{\veck}|^2\delta(\omega - \varepsilon_\veck)$, with 
$\varepsilon_\veck$ the band dispersion. The complementary asymmetric band, 
on the other hand, is decoupled from the QD, and contributes only a 
constant energy to the Hamiltonian which can be neglected.

The local Green's function at the QD site is given by the equation of 
motion
\begin{equation}
\gf{ c_{0,s}}{c_{0,s^\prime}^\+ }{\e}=\delta_{ss^\prime}+\gf{
\comm{c_{0,s}}{H}}{c_{0,s^\prime}^\+}{\e}.
\end{equation}
Evaluating the commutator
\begin{equation}
\begin{split}
\comm{c_{0,s}}{H}=& \e_{0,s}c_{0,s}+U\,n_{0,\bar
s}c_{0,s}\\ &-t_0c_{1,s}+\sum_{\veck}\tilde V_\veck c_{\veck,s},
\end{split}
\end{equation}
we obtain
\begin{equation}\label{G_00}
\begin{split}
&\left(\e-\e_{0,s}\right)\gf{ c_{0,s}}{c_{0,s^\prime}^\+
}{\e}=\delta_{ss^\prime}+U\gf{ n_{0,\bar
s}c_{0,s}}{c_{0,s^\prime}^\+}{\e}\\
&-t_0\gf{ c_{1,s}}{c_{0,s^\prime}
^\+}{\e} -\sum_\veck \tilde V_{\veck} \gf{
c_{\veck,s}}{c_{0,s^\prime}^\+}{\e}.
\end{split}
\end{equation}
The three new correlation functions on the right--hand side must be evaluated as well.
The first and last obey the equation of
motion
\begin{widetext}
\begin{equation}
\begin{split}
 &\left(\e -\e_{0,s}-U\right)\gf{ n_{0,\bar
s}c_{0,s}}{c_{0,s^\prime}^\+}{\e}=\langle n_{0,\bar s}
\rangle\delta_{ss\prime}+\langle c^\+_{0,\bar
s}c_{0,s}\rangle\delta_{s^\prime,\bar s}+\sum_\veck \tilde
V_{\veck}\gf{  n_{0,\bar s}c_{\veck,s}}{c^\+_{0,s^\prime}}{\e}
+\sum_{\veck} \tilde V_\veck\gf{ c^\+_{0,\bar s}c_{\veck,\bar s}
c_{0,s}}{c^\+_{0,s^\prime}}{\e} \\
&-\sum_{\veck}\tilde V^*_\veck\gf{ c^\+_{\veck,\bar s}c_{0,\bar s}
c_{0,s}}{c^\+_{0,s^\prime}}{\e} -t_0\gf{ n_{0,\bar s}c_{1,
s}}{c^\+_{0,s^\prime}}{\e} -t_0\gf{ c^\+_{0,\bar s}c_{1,\bar s}
c_{0,s}}{c^\+_{0,s^\prime}}{\e}+t_0\gf{ c^\+_{1,\bar s}c_{0,\bar s}
c_{0,s}}{c^\+_{0,s^\prime}}{\e}.
\end{split}
\end{equation}
\end{widetext}

At this point we use the Hubbard I decoupling procedure, introducing the
following approximations:
\begin{subequations}\label{eq:h1procedure}
 \begin{equation}
 \gf{ n_{0,\bar s}c_{1, s}}{c^\+_{0,s^\prime}}{\e}\approx
\langle n_{0,\bar s} \rangle\gf{ c_{1, s}}{c^\+_{0,s^\prime}}{\e},
\end{equation}
\begin{equation}
\begin{split}
&\sum_\veck \tilde  V_{\veck}\gf{
n_{0,\bar s}c_{\veck,s}}{c^\+_{0,s^\prime}}{\e}\\ &\quad \approx
\langle n_{0,\bar s} \rangle\sum_\veck \tilde  V_{\veck}\gf{
n_{0,\bar s}c_{\veck,s}}{c^\+_{0,s^\prime}}{\e},
\end{split}
\end{equation}
\begin{equation}
\begin{split}
 &\sum_{\veck}\tilde V_\veck\gf{ c^\+_{0,\bar s}c_{\veck,\bar s}
c_{0,s}}{c^\+_{0,s^\prime}}{\e}\\ &\quad \approx\sum_{\veck}\tilde
V_\veck\langle  c^\+_{0,\bar s}c_{\veck,\bar s}\rangle \gf{
c_{0,s}}{c^\+_{0,s^\prime}}{\e},
\end{split}
\end{equation}
\begin{equation}
\begin{split}
&\sum_{\veck}\tilde V^*_\veck\gf{ c^\+_{\veck,\bar s}c_{0,\bar s}
c_{0,s}}{c^\+_{0,s^\prime}}{\e}\\ &\quad \approx \sum_{\veck}\tilde V^*_\veck
\langle c^\+_{\veck,\bar s}c_{0,\bar s}  \rangle\gf{
c_{0,s}}{c^\+_{0,s^\prime}}{\e},
\end{split}
\end{equation}
\begin{equation}
\gf{ c^\+_{0,\bar s}c_{1,\bar s}
c_{0,s}}{c^\+_{0,s^\prime}}{\e} \approx \langle c^\+_{0,\bar
s}c_{1,\bar s} \rangle \gf{
c_{0,s}}{c^\+_{0,s^\prime}}{\e},
\end{equation}
\begin{equation}
\gf{ c^\+_{1,\bar s}c_{0,\bar s} c_{0,s}}{c^\+_{0,s^\prime}}{\e}
\approx \langle c^\+_{1,\bar s}c_{0,\bar s} \rangle\gf{
c_{0,s}}{c^\+_{0,s^\prime}}{\e}.
\end{equation}
\end{subequations}
\\
Moreover, we assume that
$\sum_{\veck}\tilde V^*_\veck \langle c^\+_{\veck,\bar s}c_{0,\bar s}
\rangle\!=\!\sum_{\veck}\tilde V_\veck\langle  c^\+_{0,\bar
s}c_{\veck,\bar s}\rangle$, and that $\langle c^\+_{0,\bar
s}c_{1,\bar s} \rangle\!=\!\langle c^\+_{1,\bar
s}c_{0,\bar s} \rangle $. With these assumptions we get
\begin{equation}\label{Gamma_00}
\begin{split}
 &\left(\e -\e_{0,s}-U\right)\gf {n_{0,\bar s}c_{0,s}}{c_{0,s^\prime}^\+}{\e}=\\
&\quad \langle n_{0,\bar s}\rangle\delta_{ss^\prime}+\langle c^\+_{0,\bar 
s}c_{0,s} \rangle\delta_{s^\prime,\bar s} +\langle n_{0,\bar s} 
\rangle\sum_\veck \tilde  V_{\veck}\gf{c_{\veck,s}} 
{c^\+_{0,s^\prime}}{\e}\\
&\quad -t_0\langle n_{0,\bar s} \rangle\gf{ c_{1,s}}{c^\+_{0,s^\prime}}{\e}.
\end{split}
\end{equation}
It is now straightforward to obtain the following expression for $\gf{
c_{\veck,s}}{c^\+_{0,s^\prime}}{\e}$:
\begin{eqnarray}\label{Gk0}
 \gf{
c_{\veck,s}}{c^\+_{0,s^\prime}}{\e}=-\frac{\tilde V_\veck^*}{\e-\e_\veck}
\gf{ c_{0 ,s}}{c^\+_{0,s^\prime}}{\e}.
\end{eqnarray}

Substituting Eq.~\eqref{Gk0} into Eq.~\eqref{Gamma_00}, and then the  
resulting expression into \eqref{G_00}, we obtain

\begin{widetext}
\begin{equation}\label{G_00-2}
\begin{split}
\left[\e-\e_{0,s}-\left(1 + \frac{U\langle
 n_{0,\bar s} \rangle}{\e-\e_{0,s}-U} \right)\sum_{\veck}
\frac{|\tilde V_\veck|^2}{\e-\e_\veck}\right]\gf{
c_{0,s}}{c_{0,s^\prime}^\+
}{\e}=& \,\, \delta_{ss^\prime}
+\frac{U\langle  n_{0,\bar s} \rangle\delta_{ss^\prime}
+U \langle c^\+_{0,\bar s}c_{0,s}\rangle\delta_{\bar ss^\prime}
}{\e-\e_{0,s}-U }\\
&-\left(1 + \frac{U\langle
 n_{0,\bar s} \rangle}{\e-\e_{0,s}-U} \right)t_0\gf{
c_{1,s}}{c^\+_{0,s^\prime}}{\e}.
\end{split}
\end{equation}
\end{widetext}

The expression above is simplified in the wide--band limit for
the electronic band, in which case $\sum_{\veck} |\tilde
V_\veck|^2/(\e-\e_\veck) =-i\,\Gamma(\e)$. We can simplify this even  
further by assuming a ``flat'' density of states, so that $\Gamma(\e) 
\equiv \Gamma$ is a constant. 
 
Some algebraic manipulation leads to the
compact form
\begin{equation}
\begin{split}
\gf{ c_{0,s}}{c_{0,s^\prime}^\+ }{\e}=&\,\tilde
g_{0s,0s}(\e)\delta_{ss^\prime}+\frac{ A_{g,s}(\e)U \langle c^\+_{0,\bar
s}c_{0,s}\rangle\delta_{\bar ss^\prime}}{(\e-\e_{0,s})(\e-\e_{0,s}-U)
} \\ &-\tilde g_{0s,0s}(\e) t_0\gf{
c_{1,s}}{c^\+_{0,s^\prime}}{\e},
\end{split}
\end{equation}
with
\begin{equation}
	A_{g,s}(\e)=\frac{1}{1+i\Gamma g_{0s,0s}(\e)},
\end{equation}
and  
\begin{eqnarray}\label{g_atomic}
g_{0s,0s}(\e)=\frac{1-\langle n_{0,\bar
s}\rangle}{\e-\e_{0,s}}+\frac{\langle n_{0,\bar s}\rangle}{\e-\e_{0,s}-U}.
\end{eqnarray}
Eqution \eqref{g_atomic} is the exact Green's function for the QD in
the atomic limit
($\tilde{V}_\veck=t_0=0$).

We now need to evaluate the Green's function $\gf{
c^\dagger_{0,s}}{c^\dagger_{0,s^\prime}}{\e}$.  We have

%
\begin{equation}\label{G_00-4}
\begin{split}
\gf{ c^\+_{0,s}}{c_{0,s^\prime}^\+
}{\e}=&\,\frac{-A_{h,s}(\e)U \langle c^\+_{0,\bar
s}c^\+_{0,s}\rangle\delta_{s^\prime\bar s} }{(\e+\e_{0,s})(\e+\e_{0,s}+U)
}\\ &+\tilde h_{0s,0s} t_0\gf{  c^\+_{1,s}}{c^\+_{0,s^\prime}}{\e},
\end{split}
 \end{equation}
where
\begin{equation}
	\tilde h_{0s,0s}(\e)= A_{h,s}(\e)h_{0s,0s}(\e),
\end{equation}
\begin{equation}
	A_{h,s}(\e)=\frac{1}{1+i\Gamma h_{0s,0s}(\e)},
\end{equation}
and
\begin{equation}
h_{0s,0s}(\e)=\frac{1+\langle n_{0,\bar
s}\rangle}{\e+\e_{0,s}}-\frac{\langle n_{0,\bar s}\rangle}{\e+\e_{0,s}+U}.
\end{equation}

Finally, we can write the Green's function for the QD in the limit of 
$t_0=0$,  within the Hubbard I approximation, as
\begin{widetext}
\begin{equation}\label{g_dot}
{\bf g}_{0,0}(\e)=
\begin{bmatrix}
\tilde g_{0\up,0\up}(\e)& \frac{A_{g,\uparrow}(\e)U \langle c^\+_{0,\dn}
c_{0,\up}\rangle}{(\e- \e_{0,\up})(\e-\e_{0,\up}-U)
} &0&\frac{A_{g,\uparrow}(\e)U \langle c_{0,\dn}c_{0,\up}\rangle}{(\e-
\e_{0,\up})(\e-\e_{0,\up}-U) }\\
\frac{A_{g,\downarrow}(\e)U \langle
c^\+_{0,\up}c_{0,\dn}\rangle}{(\e-\e_{0,\dn})(\e-\e_{0,\dn}-U)
} &\tilde g_{0\dn,0\dn}(\e)&
\frac{A_{g,\downarrow}(\e)U \langle c_{0,\up}c_{0,\dn}\rangle}{(\e-
\e_{0,\dn})(\e-\e_{0,\dn}-U) }
&0\\
0&\frac{A_{h,\uparrow}(\e)U \langle c^\+_{0,\up}c^\+_{0,\dn}\rangle}
{(\e+\e_{0,\up})(\e+\e_{0,\up}+U) }&\tilde h_{0\up,0\up}(\e)&
\frac{A_{h,\uparrow}(\e)U \langle c_{0,\dn} c^\+_{0,\up}\rangle}{(\e+\e_{0,\up})
(\e+\e_{0,\up}+U)}
\\
\frac{A_{h,\downarrow}(\e)U \langle c^\+_{0,\dn}c^\+_{0,\up}\rangle}{(\e+
\e_{0,\dn})(\e+\e_{0,\dn}+U) }&0&\frac{A_{h,\downarrow}(\e)U \langle
c_{0,\up} c^\+_{0,\dn}\rangle}{(\e+\e_{0,\dn})(\e+\e_{0,\dn}+U)
}&\tilde h_{0\dn,0\dn}(\e)
\end{bmatrix}.
\end{equation}
\end{widetext}
 The coupling of the QD with the first site of the wire is given
by the matrix
\begin{eqnarray}
 {\bf t}_0=
\begin{pmatrix}
-t_0&0& 0 &0\\
0&-t_0&0&0\\
0&0&t_0&0\\
0&0&0&t_0
\end{pmatrix}.
\end{eqnarray}
Note that the Green's function matrix \eqref{g_dot}  depends  on
various expectation values. The two occupations $\langle
n_{0,\up}\rangle$ and $\langle
n_{0,\dn}\rangle$, appearing in the diagonal elements of
${\bf g}_{0,0}(\e)$, are given at finite temperature by
\begin{eqnarray}
 \langle n_{0,\up}\rangle=-\frac{1}{\pi}\int_{-\infty}^\infty \ud \e  
\,f(\e,\,T) {\rm Im
}~[{\bf  G}_{0,0}(\e)]_{1,1},
\end{eqnarray}
and
\begin{eqnarray}
 \langle n_{0,\dn}\rangle=-\frac{1}{\pi}\int_{-\infty}^\infty \ud \e  
\,f(\e,\,T) {\rm Im
}~[{\bf  G}_{0,0}(\e)]_{2,2},
\end{eqnarray}
where $f(\e,\,T)=\left[1+\exp\left(\e/T \right) \right]^{-1}$ is the Fermi function.
There are eight other expectation values appearing in the off--diagonal
terms of the matrix \eqref{g_dot}. However, due to the
anticommutation relations between fermionic operators, only four of them  
are independent:
\begin{subequations}
\begin{equation}
\begin{split}
\langle c^\+_{0,\dn} c_{0,\up}\rangle &=-\langle c_{0,\up} c^\+_{0,\dn}
\rangle\\ &=-\frac{1}{\pi}\int_{-\infty}^\infty \ud \e\, f(\e,\,T) {\rm Im }~[{\bf
G}_{0,0}(\e)]_{1,2},
\end{split}
\end{equation}
\begin{equation}
\begin{split}
\langle c^\+_{0,\up} c_{0,\dn}\rangle &=-\langle c_{0,\dn} c^\+_{0,\up}
\rangle\\
&=-\frac{1}{\pi}\int_{-\infty}^\infty \ud \e\, f(\e,\,T) {\rm Im }~[{\bf
G}_{0,0}(\e)]_{2,1},
\end{split}
\end{equation}

\begin{equation}
\begin{split}
\langle  c_{0,\dn} c_{0,\up}\rangle &=-\langle c_{0,\up} c_{0,\dn} \rangle\\
&=-\frac{1}{\pi}\int_{-\infty}^\infty \ud \e\, f(\e,\,T) {\rm Im }~[{\bf
G}_{0,0}(\e)]_{1,4},
\end{split}
\end{equation}
and
\begin{equation}
\begin{split}
\langle  c^\+_{0,\up} c^\+_{0,\dn}\rangle &=-\langle c^\+_{0,\dn}
c^\+_{0,\up} \rangle \\
&=-\frac{1}{\pi}\int_{-\infty}^\infty \ud \e\,
f(\e,\,T) {\rm Im }~[{\bf  G}_{0,0}(\e)]_{4,1}.
\end{split}
\end{equation}
\end{subequations}

These expectation values depend on the Green's functions themselves, and  
thus have to be computed self--consistently. 

\section{Charge and spin polarization of the gated quantum dot coupled to a 
Majorana mode} \label{app:splitting}
In this appendix  we justify our interpretation of the results presented in 
Fig.\ \ref{fig:KondoQuenchVg}, Sec.\ \ref{sec:experimental}. Equation 
(\ref{eff_Zeeman}) gives the Majorana--induced effective Zeeman splitting 
$V_Z^{(M)}$ for a small detuning from particle--hole symmetry, $|V_g| \ll 
|\varepsilon_{\text{dot}}|$.

\begin{figure}
\begin{center}
\includegraphics[width=0.95\columnwidth]{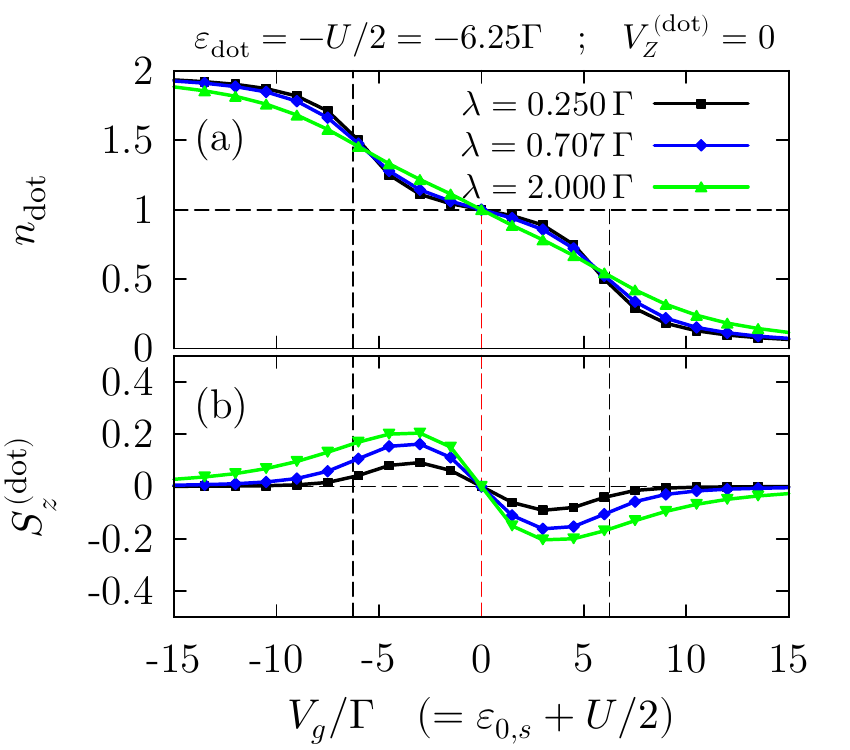}
\caption{(Color online) Quantum dot occupancy (charge) and spin 
polarization for  the parameters of Fig.\ 
\ref{fig:KondoQuenchVg}.\label{fig:splitting}}
\end{center}
\end{figure}

For large enough $\lambda$, a small $V_g$ will quench the Kondo effect by 
breaking  the spin symmetry. For $\lambda = 2\,\Gamma$ (triangles), Figure 
\ref{fig:splitting}(a) shows a rapid change in charge for both positive and 
negative $V_g$. For $|V_g|$ as small as $2.5\,\Gamma$, Fig.\ 
\ref{fig:KondoQuenchVg}(c) demonstrates that the conductance enhancement 
due to the Kondo effect has decreased as much as $50\%$, even though 
the system is far from the mixed--valence regime. The negative (positive) 
spin polarization of the QD for positive (negative) $V_g$ shown in Fig.\ 
\ref{fig:splitting}(b) demonstrates it is a Zeeman splitting accompanying 
the level shift by the gate voltage that kills the Kondo effect.

On the other hand, for small enough $\lambda$ the Majorana--induced  Zeeman 
splitting may be weak enough for the Kondo effect to survive until the 
mixed valence regime is reached, at which point it is quenched by charge 
fluctuations within the QD. For $\lambda = 0.25\,\Gamma$ (squares), Fig.\ 
\ref{fig:KondoQuenchVg}(c) shows a much slower decay of the enhanced 
conductance, where only for $|V_g| \sim |\varepsilon_{\text{dot}}|$ has the 
conductance contribution from the Kondo effect been reduced by half. At 
this point, indicated in Fig.\ \ref{fig:splitting} by the lateral vertical 
dashed lines, Fig.\ \ref{fig:splitting}(a) shows that the QD is far away 
from single occupancy, this time with a negligible spin polarization 
presented in Fig.\ \ref{fig:splitting}(b). This is clear indication that the mixed--valence 
regime has been reached, and the Kondo effect finally disappears due to 
charge fluctuations in the QD.


\end{document}